\pdfoutput=1
\documentclass[10pt]{article}

\usepackage[a4paper, margin=2.5cm]{geometry}
\usepackage[british]{babel}
\usepackage{csquotes}
\usepackage{graphicx}
\usepackage{float}
\usepackage[hidelinks]{hyperref}
\usepackage{booktabs}
\usepackage{tabularx}
\usepackage{amsmath}
\usepackage{subcaption}
\usepackage{array}
\usepackage{threeparttable}
\usepackage{parskip}
\usepackage{multirow}
\usepackage{tcolorbox}
\tcbuselibrary{breakable}
\usepackage{enumitem}
\usepackage{newtxtext}
\usepackage{newtxmath}

\newtcolorbox{conversation}{
    breakable,
    colback=black!2,
    colframe=black!20,
    boxrule=0.6pt,
    arc=0pt,
    left=12pt,
    right=12pt,
    top=9pt,
    bottom=9pt,
    toprule at break=0pt,
    bottomrule at break=0pt
}

\usepackage[
  style=apa,
  backend=biber,
  natbib=true,
  doi=true,
  url=true
]{biblatex}  % APA 7th bibliography

\graphicspath{ {images/} }

\hypersetup{
  pdftitle={From Dialogue to Task: How Cognitive Engagement With Generative Artificial Intelligence Relates to Post-Dialogue Activity and Task Outcomes},
  pdfauthor={Xiaoyu Zhuang, Aneesha Bakharia, Dragan Gašević, Hassan Khosravi}
}

\begin{document}

% ---- Title block (custom, replaces \maketitle for tighter spacing) ----
\begin{center}
  \vspace*{-1.5em}
  {\LARGE From Dialogue to Task: How Cognitive Engagement With Generative Artificial Intelligence Relates to Post-Dialogue Activity and Task Outcomes\par}
  \vspace{1.5em}
  {\large
    Xiaoyu Zhuang\textsuperscript{1,*}\qquad
    Aneesha Bakharia\textsuperscript{1}\qquad
    Dragan Ga\v{s}evi\'{c}\textsuperscript{2,3}\qquad
    Hassan Khosravi\textsuperscript{4}\par}
  \vspace{0.9em}
  {\small
    \textsuperscript{1}School of Electrical Engineering and Computer Science, The University of Queensland, Australia\\
    \textsuperscript{2}Centre for Learning Analytics, Faculty of Information Technology, Monash University, Australia\\
    \textsuperscript{3}Faculty of Education and School of Computing and Data Science, The University of Hong Kong, Hong Kong, China\\
    \textsuperscript{4}Institute for Teaching and Learning Innovation, The University of Queensland, Australia\\[0.4em]
    \textsuperscript{*}Corresponding author: \href{mailto:x.zhuang@uq.edu.au}{x.zhuang@uq.edu.au}\par}
  \vspace{1.2em}
  \parbox{0.82\textwidth}{\centering\small\itshape
    Preprint. This manuscript has been submitted to the Journal of Computer Assisted Learning and is currently under review. It has not been peer reviewed.}
\end{center}
\vspace{0.4em}

%TC:ignore
% Front Matter
\begin{abstract}
\noindent\textbf{Background:} Generative AI is increasingly embedded within learning tasks, yet interaction frequency reveals little about cognitive engagement in student-AI dialogue. Little is known about how such engagement is situated within task workflows or how dialogue characteristics and subsequent task activity relate to students' behaviour and task outcomes.

\noindent\textbf{Objectives:} This study examined cognitive engagement in student-AI dialogue, its position within task workflows, and associations of dialogue and task-process characteristics with post-dialogue activity and task outcomes.

\noindent\textbf{Methods:} Data were collected from 1,681 students across 40 university courses using three AI companions embedded in resource creation, peer moderation, and practice. Cognitive engagement in 829 conversation threads was coded with a dialogue-oriented Interactive-Constructive-Active-Passive (ICAP) framework, and 2,969 task sequences were modelled with first-order Markov models. Regression models linked dialogue and task-process characteristics to behavioural re-engagement, peer-rated artefact quality, and practice performance.

\noindent\textbf{Results and Conclusions:} Passive and Active engagement predominated, although resource creation involved more Constructive and Interactive dialogue, peer moderation more Passive, and practice more Active, while higher-engagement dialogue was less often followed by session termination. In resource creation, student-composed messages were associated with more dialogue-to-task returns than template-only requests at the same Passive level, whereas deeper cognitive engagement was associated with higher final artefact quality independent of return frequency; in practice, AI dialogue after an incorrect attempt was associated with fewer further attempts but higher next-attempt correctness among continuers. Post-dialogue activity and task outcomes are distinct aspects of AI use that interaction frequency cannot capture and evaluations should assess separately.
\end{abstract}


%TC:endignore

% Main Text
\section{Introduction}
Rapid advances in large language models (LLMs) \citep{zhao2026survey} have accelerated the integration of artificial intelligence (AI) into higher education \citep{wang2024aied_lr}, where generative AI (GenAI) tools increasingly provide tutoring, personalised feedback, and on-demand cognitive support for learners \citep{heung2025engagement, qian2025pedagogical, giannakos2025genai_ed, wu2025genai_review}. Early empirical studies of this integration highlighted that students often interacted with GenAI in an isolated, dyadic manner outside the classroom \citep{park2024promise_peril_chatgpt_he}, leaving AI interaction analytically separated from the task itself. More recent learning platforms embed GenAI directly within authoring, evaluation, and practice activities, supplying the model with current task content and making AI support part of students' ongoing coursework \citep{pozdniakov2025cocreation}. This shift directs attention to how students engage with AI while completing specific, authentic learning tasks, beyond whether and how often they access it \citep{uzun2026la_feedback}.

Educational dialogue may support learning by enabling learners to articulate ideas, reason, and co-construct understanding \citep{alexander2008dialogic_teaching, cui2023dialogue}. Student-AI dialogue has become an increasingly prevalent form of educational interaction, offering readily available and responsive conversational support \citep{debets2025review, park2024promise_peril_chatgpt_he}. Yet interaction volume provides limited information about what this dialogue may accomplish cognitively \citep{weatherholtz2025cognitive}. A high volume of interaction may reflect dynamic and iterative reasoning \citep{bai2025impact_genai}, but can also mask repeated answer-seeking \citep{zhai2024effects} or cognitive offloading and passive delegation to AI-generated output \citep{fan2025laziness}. This has renewed interest in frameworks that distinguish different forms of students' cognitive engagement \citep{heung2025engagement, sakeef2025detecting}. ICAP is a well-known framework in the learning sciences and offers one such lens by differentiating Passive, Active, Constructive, and Interactive contributions \citep{chi2014icap}, and has begun to be adapted to GenAI tutoring conversations at scale \citep{weatherholtz2025cognitive}. However, ICAP was developed for classroom and human-to-human discourse \citep{chi2014icap, vosniadou2023icap_cg}, in which Interactive engagement requires both partners to contribute generatively and to build on each other's contributions over sufficient turns \citep{chi2014icap}. When the partner is an AI that can produce elaborate content on demand, co-construction cannot be inferred from exchange frequency itself, but only from how the student takes up and builds on AI responses across multiple turns. Temporal analyses of human-to-human online discussion have shown that cognitive engagement develops as interaction unfolds \citep{guo2023understanding}. In contrast, analyses of student-AI dialogue still largely code individual utterances or turns, leaving the development of cognitive engagement across a conversation comparatively underexamined \citep{zheng2025rely}.

Research on student-AI interaction commonly examines dialogue content and process traces separately, leaving only a partial account of how AI use unfolds within learning tasks. Coding conversation content can reveal what students contribute cognitively, but because such coding typically treats messages in isolation, it cannot show when cognitive contributions occur within the task or what students do with them afterwards \citep{zheng2025rely}. Conversely, platform logs of authoring, feedback, revision, practice, and submission can offer insights into the sequence of task activity \citep{pozdniakov2025cocreation}, but provide limited insight into the metacognitive and inferential processes represented within each AI interaction \citep{weatherholtz2025cognitive}. Because AI use on a learning platform unfolds within a temporal workflow \citep{pozdniakov2025cocreation}, interpreting a dialogue requires attention to both what the student contributes and where the dialogue occurs in the task sequence \citep{chen2018temporal, knight2017temporal}. Analysing dialogue content and process traces separately also limits explanations of task outcomes, because similar levels of AI use can involve different forms of cognitive engagement \citep{uzun2026la_feedback} and different patterns of subsequent task activity \citep{fan2025laziness}. Consequently, despite growing interest in learning analytics for GenAI-supported learning, it remains unclear how the cognitive engagement evident in student-AI dialogue relates to where that dialogue occurs within a task, what students do after it, and the task outcomes that follow \citep{misiejuk2025mapping, yan2024la_cycle}. Addressing this requires combining process traces with discourse-level analysis of cognitive engagement \citep{alamray2026explore, chen2025unpacking}.

The present study examined how students cognitively engaged with task-embedded AI, how these interactions were situated within ongoing task processes, and how dialogue and process characteristics related to post-dialogue activity and task outcomes. The study was conducted on an authentic learning platform where three task-embedded AI companions support students as they author learning resources, evaluate peers' work, and practise disciplinary concepts across multiple university courses. We classified cognitive engagement in complete conversation threads using a dialogue-oriented adaptation of the ICAP framework, modelled the task sequences in which these threads were embedded using first-order Markov models, and connected these characteristics to post-dialogue activity and task outcomes. In doing so, the study distinguishes what students contribute within AI dialogue from where that dialogue occurs in the task process and what follows it, allowing these dimensions to be examined as related but non-equivalent aspects of task-embedded AI use. Three research questions guided the study:
\begin{enumerate}[label=\textbf{RQ\arabic*.}, leftmargin=*]
    \item \textbf{Cognitive Engagement With AI:} What forms of cognitive engagement characterise students' dialogue with task-embedded AI, and how do they vary across task contexts?
    \item \textbf{Process Patterns of AI Use:} Where is student-AI dialogue situated within task workflows, and what task-process patterns are associated with different forms of cognitive engagement?
    \item \textbf{Post-Dialogue Activity and Task Outcomes:} To what extent are dialogue and task-process characteristics associated with behavioural re-engagement, artefact quality, and immediate practice performance?
\end{enumerate}

\section{Literature Review}
\subsection{Task-Embedded GenAI-Supported Learning}
GenAI can support tutoring, formative feedback, ideation, and content generation in higher education \citep{giannakos2025genai_ed, qian2025pedagogical, wu2025genai_review}, but its educational value depends less on raw model capability than on how it is woven into curricular activities, task requirements, and feedback loops \citep{cukurova2025interplay, weidlich2025chatgpt}. Earlier research frequently examined conversational AI as a standalone resource separate from students' coursework platforms, leaving interactions analytically detached from the learning task and reliant on self-reported surveys or user evaluations \citep{debets2025review, luo2025design_ai_tools_lr, park2024promise_peril_chatgpt_he}. To capture task-centric learning dynamics, subsequent system designs began embedding conversational agents directly alongside learning materials to track structured interaction logs \citep{schmucker2024ruffle_riley}. More recent platform-embedded designs instead surface AI support during authoring, evaluation, and revision, supplying the model with the current task content and stage and allowing AI interaction to be linked directly to what students subsequently do with their work \citep{alsaiari2025feedback, pozdniakov2025cocreation}.

Embedding GenAI within learning tasks changes what a single AI conversation represents. The same request for help can serve as feedback-seeking on a draft during authoring, scrutiny of a peer's resource during moderation, or concept clarification during practice, depending on the task stage, the assigned companion role, and interface affordances, such as predefined prompt suggestions \citep{alamray2026explore}, that shape how a request is phrased in the first place \citep{alsaiari2025feedback, lee2025prompt, pozdniakov2025cocreation}. Therefore, understanding task-embedded GenAI requires attention to both the cognitive engagement expressed in student-AI dialogue \citep{weatherholtz2025cognitive, yang2025analysing} and where that dialogue occurs in the sequence of task activity and what students do after it \citep{chen2025unpacking, fan2025laziness}.

\subsection{Cognitive Engagement in Student-AI Dialogue}
Engagement is commonly conceptualised across behavioural, emotional, and cognitive dimensions \citep{fredricks2004school_engagement}. Among these, cognitive engagement, characterised as psychological investment and thoughtful effort directed towards understanding complex ideas, is widely recognised as essential for higher-order learning and conceptual transfer \citep{chi2014icap, fredricks2004school_engagement}. However, much research on GenAI in education continues to operationalise student engagement through broad outcome measures or superficial proxies, such as frequency of interactions, user satisfaction, or self-reported motivation \citep{alamray2026explore, debets2025review, wu2025genai_review}. Such metrics fail to capture the qualitative depth of cognition, as they cannot distinguish a student who repeatedly re-requests an answer from one whose concise contribution reflects deliberate reasoning, nor does an elaborate AI output guarantee deep student processing \citep{alamray2026explore, chen2025unpacking, zheng2025rely}. Therefore, a more theoretically grounded account requires a framework that differentiates forms of cognitive engagement based on learners' observable contributions. The ICAP framework, proposed by \citet{chi2014icap} and widely known in the learning sciences, provides such a conceptualisation by categorising overt learner behaviours into Passive, Active, Constructive, and Interactive modes, which correspond to progressively deeper cognitive processes. Applied to authentic classroom observations, ICAP has demonstrated that instructional time is predominantly occupied by Passive and Active behaviours, leaving only a small fraction for Constructive or Interactive engagement \citep{vosniadou2023icap_cg}.

For student-AI dialogue, ICAP offers a compelling conceptual lens because it shifts attention from prompt frequency towards the overt forms of cognitive processing expressed in student contributions \citep{chi2018practice, weatherholtz2025cognitive}. Empirical work has initiated this extension across diverse educational contexts: \citet{gao2024icap_genai} adapted ICAP to evaluate ChatGPT integration and its relation to student motivation, \citet{chen2025genai_icap} operationalised ICAP to analyse pre-service teachers' instructional-design dialogues, and \citet{weatherholtz2025cognitive} validated a scalable LLM-as-judge procedure to classify cognitive engagement in automated math tutoring. Nevertheless, while earlier studies relied on self-report survey scales \citep{gao2024icap_genai} or specialised teacher cohorts \citep{chen2025genai_icap} and recent automated implementations remain tailored to structured tutoring \citep{weatherholtz2025cognitive}, there remains a lack of empirical clarity on how cognitive engagement unfolds during naturally occurring, task-embedded dialogues across broader and authentic learning activities \citep{chen2025unpacking, knight2017temporal, uzun2026la_feedback}.

A further challenge is that ICAP was originally formulated for traditional classroom discourse and instructional materials \citep{chi2009aci, chi2014icap, chi2018practice, chi2021gap, vosniadou2023icap_cg}, rather than iterative and adaptive AI dialogue. Directly applying ICAP to student-AI interactions introduces several adaptation problems. First, because ICAP infers cognition from observable output \citep{chi2014icap, chi2018practice}, coding schemes need to rigorously isolate student contributions from AI-generated content. An elaborately structured AI response can create an illusion of high-level cognitive activity even when the student's own contribution is minimal \citep{alamray2026explore, zheng2025rely}. Second, student-AI interactions unfold as dynamic, multi-turn conversational trajectories beyond isolated utterances \citep{alamray2026explore, shen2025process, zheng2025rely}. Consequently, evaluating Interactive or Constructive engagement mostly requires attention to complete conversation threads rather than isolated messages, as repeated question-and-answer exchanges do not inherently signify knowledge co-construction \citep{shen2025process, uzun2026la_feedback, weatherholtz2025cognitive}. Third, template prompts or platform starters may artificially inflate interaction volume without reflecting student-initiated reasoning \citep{alamray2026explore, lee2025prompt}.

\subsection{AI Use as a Task Process}
Dialogue analysis can offer insights into how students communicate with AI, but it cannot show where dialogue occurs in the task sequence or what task activity follows it without linkage to task logs. Much existing research still compresses AI use into aggregate indicators that describe overall intensity, such as usage frequency, duration, or composite engagement scores, offering limited insight into the temporal dynamics of learning \citep{alamray2026explore, knight2017temporal, liu2025assessing, uzun2026la_feedback}. Crucially, high interaction frequency does not by itself signal strategic or effective learning \citep{liu2025assessing}, and sequential patterns of learner actions can distinguish productive activity where aggregated counts of the same actions do not \citep{chen2018temporal}. Indeed, the same volume of interaction may represent distinct learning pathways depending on whether GenAI is consulted during early ideation, immediately following feedback, or as a late substitute for critical revision \citep{fan2025laziness, pozdniakov2025cocreation}.

Sequence analysis and process mining can reveal how student-AI interaction is situated within the temporal structure of task activity by treating platform logs as ordered transitions between operational states \citep{chen2025unpacking, chen2018temporal, helske2024process}. Such temporal methods have been advocated for capturing learning as a process that unfolds over time \citep{knight2017temporal, chen2018temporal} and are increasingly used to examine the implications of GenAI for human learning \citep{chen2025unpacking}. For instance, \citet{fan2025laziness} used process mining of self-regulated learning sequences to show that students consulting ChatGPT during revision formed distinct closed loops between writing and AI queries compared to those assisted by human experts. However, few studies have explicitly connected the cognitive engagement evident in student-AI dialogues with the process patterns through which tasks were actually completed \citep{alamray2026explore, misiejuk2025mapping, zheng2025rely}. First-order Markov models (FOMMs) are particularly well suited to bridging this gap. By estimating the transition probabilities between adjacent task states, they reveal dominant local transitions and recurring cycles in an interpretable form, despite capturing only immediate-step dependencies \citep{chen2025unpacking, helske2024process}. Despite this groundwork, how students integrate GenAI dialogue across the authoring, evaluating, and revising activity of authentic coursework remains largely unexamined \citep{shen2025process, yan2024la_cycle, zheng2025rely}.

\subsection{Student-AI Interaction and Task Outcomes}
Evidence on how GenAI relates to students' work and learning remains difficult to interpret as many comparison studies confound GenAI use with added feedback or scaffolding and some measure performance while students can still use the tool \citep{lawson2026media_comparisons, weidlich2025chatgpt, yan2025gains}. AI-generated feedback performs comparably to human feedback on average, despite marked variability across studies \citep{kaliisa2026ai_feedback}, and the quality of students' work has been linked to how students interact with AI rather than to use alone \citep{bai2025impact_genai}. Nevertheless, AI support does not reliably translate into deeper revision, and students may adopt AI suggestions with limited critical scrutiny \citep{chen2025unpacking, fan2025laziness, zhai2024effects, zheng2025rely}. Sustained assistance may also fail to transfer into self-regulated performance once support is withdrawn \citep{chen2025unpacking, darvishi2024impact, yan2025gains}. These findings indicate that the association between GenAI use and the quality of students' work is conditional on task design, feedback type, and how AI is woven into the workflow.

Variation in artefact quality may partly depend on how students evaluate and use AI-generated feedback during task completion. Feedback effectiveness depends on the level and function it targets \citep{hattie2007feedback}, and on students' capacity to appreciate, judge, and act on what they receive (feedback literacy) \citep{carless2018feedback, uzun2026la_feedback}. Existing work demonstrates that the relationship between AI use and the quality of students' work can be analysed across four features that should not be collapsed into a single measure of AI use: whether AI was used at all (exposure); how often it was used (intensity); whether cognitive engagement was evident in their dialogue \citep{chi2014icap, chi2018practice, chen2025genai_icap}; and where in the task sequence the dialogue occurred and what task activity followed it \citep{chen2025unpacking, pozdniakov2025cocreation, shen2025process}. Research on AI-assisted co-creation further suggests that the quality of task outcomes depends on how AI assistance is incorporated into authoring and revision \citep{pozdniakov2025cocreation}. Consequently, exposure and intensity alone are insufficient for explaining task outcomes. Analyses also need to consider the cognitive engagement evident in student-AI dialogue and the task activity surrounding it.

\section{Methodology}
\subsection{Research Design}
\begin{figure}[H]
    \centering
    \includegraphics[width=1\linewidth]{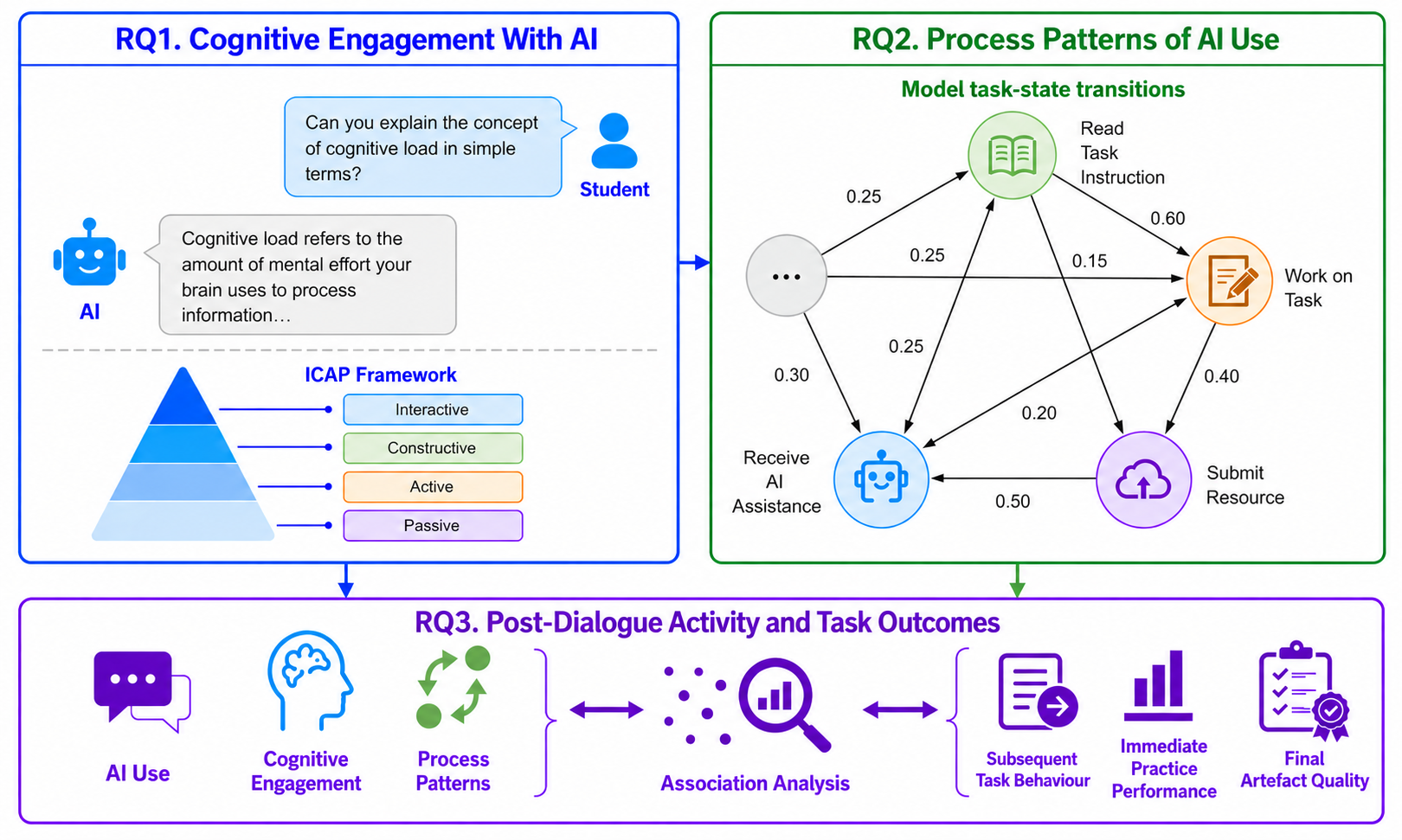}
    \caption{Overview of the research design.}
    \label{fig:research_design}
\end{figure}

This study adopted a multi-method learning analytics design (Figure~\ref{fig:research_design}), combining theory-driven qualitative coding of dialogue, sequential process modelling, and statistical modelling of task outcomes. RQ1 characterised the cognitive engagement evident in student-AI dialogue. RQ2 modelled the task-state transitions within which those conversations were embedded. RQ3 examined associations with behavioural re-engagement, final artefact quality, and immediate practice performance.

\subsection{Research Context}
\begin{figure}[H]
    \centering
    \begin{subfigure}[t]{0.49\textwidth}
        \centering
        \includegraphics[width=\linewidth]{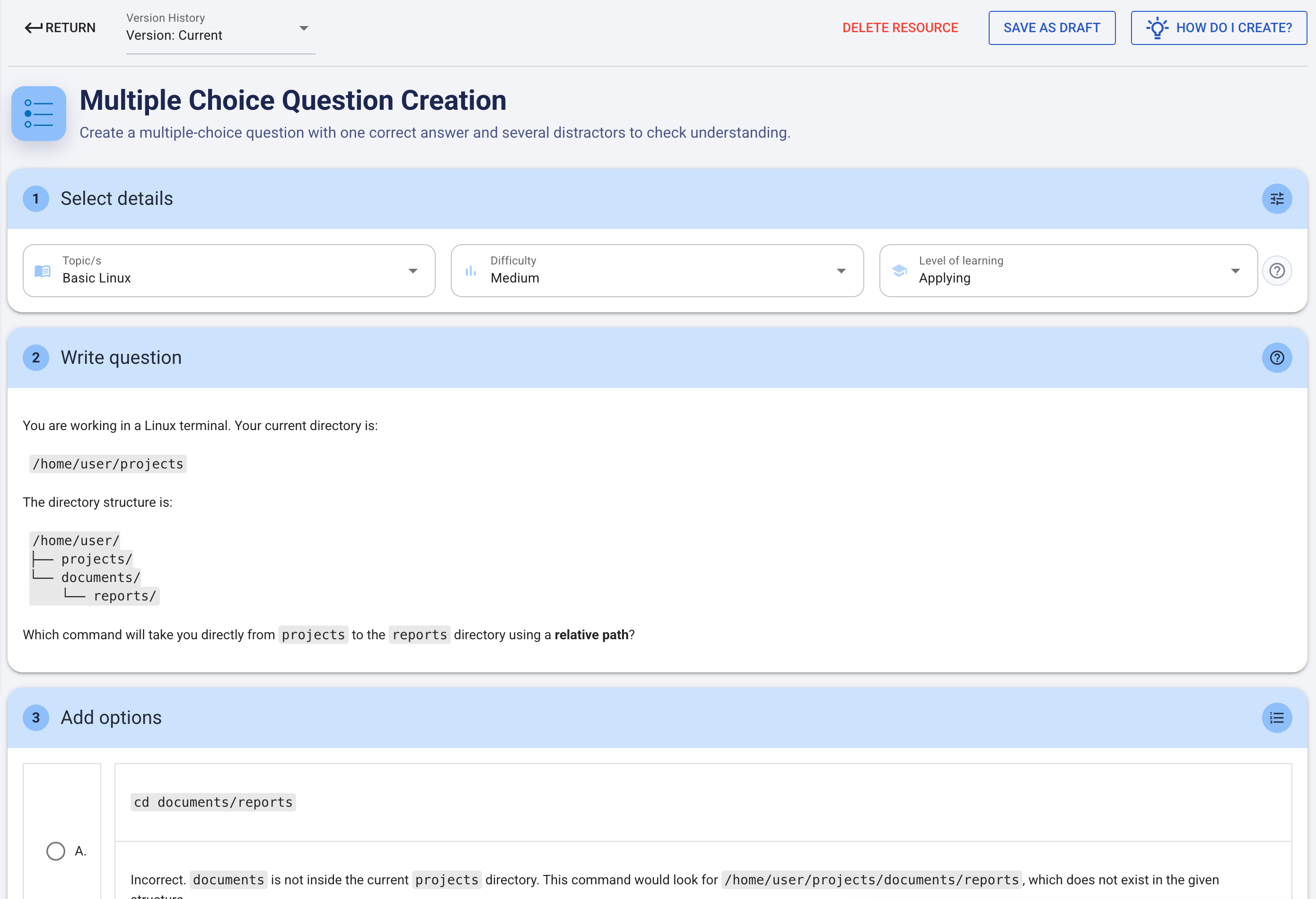}
    \end{subfigure}
    \hfill
    \begin{subfigure}[t]{0.49\textwidth}
        \centering
        \includegraphics[width=\linewidth]{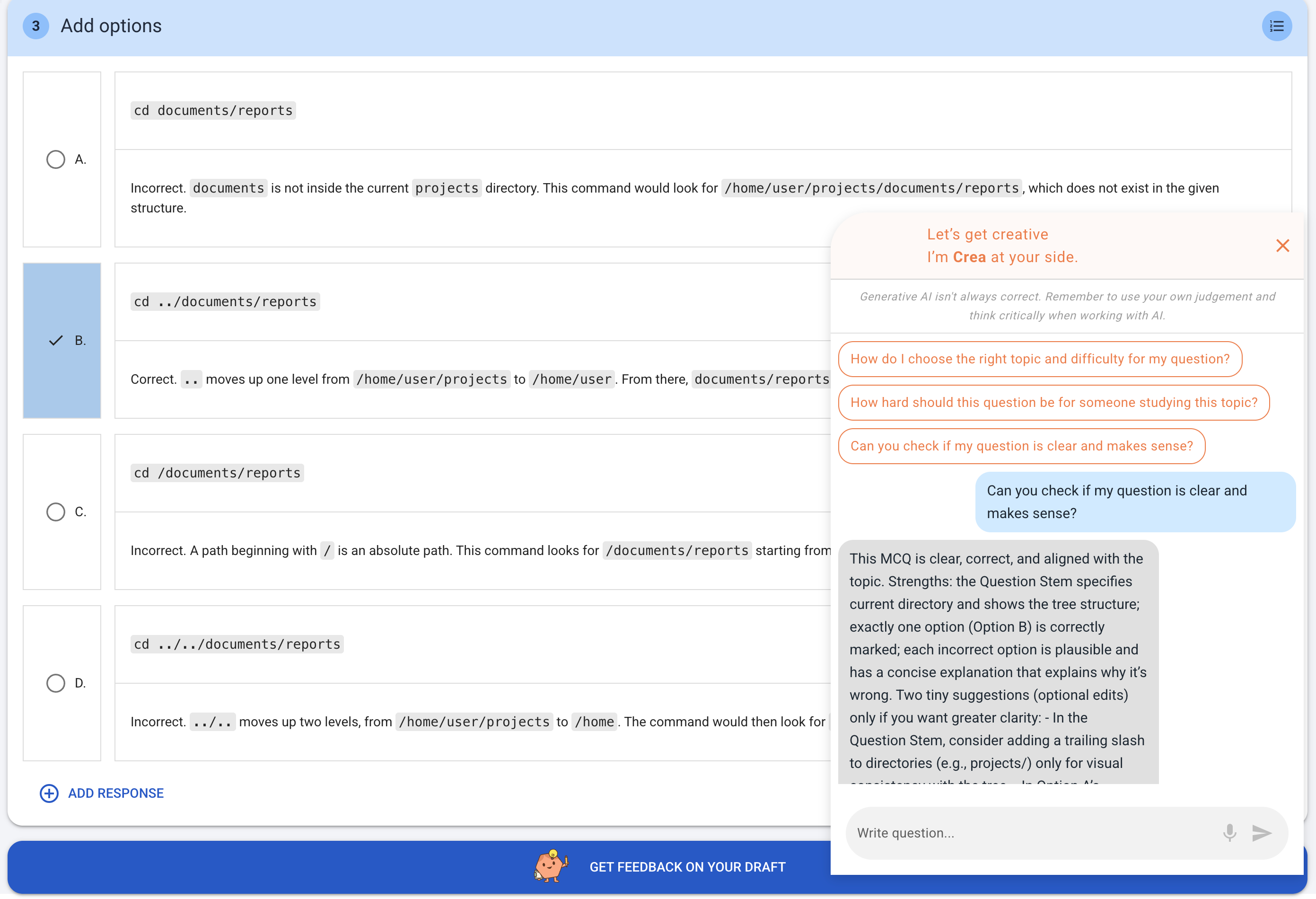}
    \end{subfigure}
    \caption{User interface for creating multiple-choice questions.}
    \label{fig:crea_ui}
\end{figure}

\begin{figure}[H]
    \centering
    \includegraphics[width=1\linewidth]{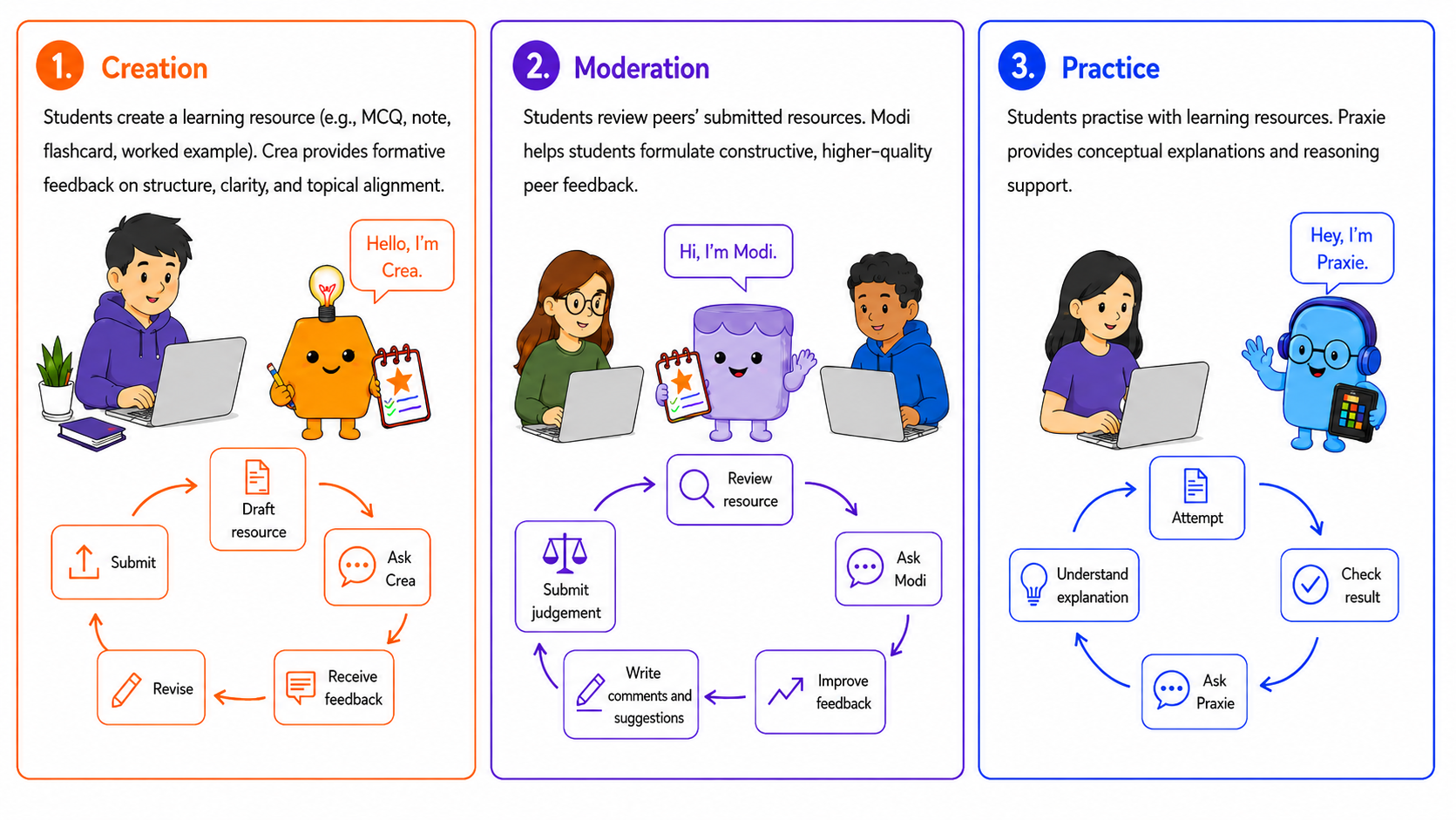}
    \caption{Three-stage AI learning workflow in RiPPLE.}
    \label{fig:learning_workflow}
\end{figure}

RiPPLE, developed by \citet{khosravi2019ripple}, is an online learning platform where students can cyclically: (1) create learning resources such as multiple-choice questions (Figure~\ref{fig:crea_ui}), notes, flashcards, and worked examples, (2) evaluate and moderate peers' resources, and (3) practise with resources to consolidate or extend understanding. As illustrated in Figure~\ref{fig:learning_workflow}, generative AI was embedded in these activities through three task-specialised conversational companions. \textit{Crea} provided formative feedback on the structure, clarity, and topical alignment of resources during Creation; \textit{Modi} supported peer moderation by helping students formulate constructive, higher-quality reviews during Moderation; and \textit{Praxie} provided conceptual explanations and reasoning support during Practice. Each companion functioned as a task-embedded scaffolding agent rather than a general-purpose chatbot. Students could open the companion at any time while working on a task and could also select conversation-starter buttons that submitted a predefined template message (Figure~\ref{fig:crea_ui}), such as a request for feedback or help choosing a resource topic. The platform recorded whether each prompt was template-generated or student-composed, a distinction retained in the outcome analysis. Each request included the current task context, such as resource type, topic, content, moderation information, or practice state. Students' interaction was thus shaped by the learning stage, the companion's assigned role, and the available starters and templates \citep{lee2025prompt}, while students could subsequently enter their own questions and engage in multi-turn dialogue.

\subsection{Data Source and Collection}
Data comprised platform logs generated during 1,681 students' routine use of RiPPLE across 40 university courses spanning computer science, medicine, business and economics, social sciences, mathematics, and languages and applied linguistics between June 2025 and June 2026. The platform automatically captured timestamped task actions and student-AI interactions (Table~\ref{tab:data_sources}), linked across user, resource, session, and conversation levels via platform-assigned identifiers. Students were informed and agreed before using the platform that platform-generated data would be collected for research purposes. Data collection and use were approved by the university's human research ethics committees. All data were anonymised before processing.

\begin{table}[H]
\centering
\caption{Data sources and descriptions.}
\label{tab:data_sources}
\begin{tabularx}{\textwidth}{p{0.28\textwidth}X}
\toprule
\textbf{Data source} & \textbf{Description} \\
\midrule

AI companion interactions &
Full chat history for each student-AI interaction across Creation, Moderation, and Practice, with the current learning context, activity stage, interaction modality, and associated resource. \\

Automated resource feedback &
System-triggered AI review requested during/after drafting, with feedback content, engagement duration, suggestions selected, and any subsequent ratings or comments. \\

Resource authoring logs &
Students' operational information that shows successive content changes and responses to structured self-evaluation prompts, reflecting how each resource was drafted, reviewed, and revised during Creation. \\

Peer moderation activities &
During Moderation, students rated submitted resources, commented on strengths and limitations, and proposed improvements. The logs retained these reviews and subsequent revisions. \\

Practice response records &
Each practice attempt recorded the response outcome, correctness, and sequence position during Practice, together with resource difficulty from learner-level and resource-level estimates. \\

\bottomrule
\end{tabularx}
\end{table}

\subsection{Analytical Framework and Dimensions}
\subsubsection{Cognitive Engagement Measurement (RQ1)}
Students' cognitive engagement in student-AI dialogue was measured using an adapted dialogue-oriented ICAP framework (Table~\ref{tab:icap_framework}) based on learners' observable contributions \citep{chi2009aci, chi2014icap}. Engagement was inferred primarily from students' own messages; AI responses were considered only to establish conversational context or evidence of student uptake. The complete conversation thread served as the unit of analysis because higher-order engagement, particularly Interactive, can emerge through sustained uptake and co-construction across multiple conversational turns, not merely within isolated messages \citep{chi2009aci, chi2014icap}. Each thread received one mutually exclusive label: Off-task (\textit{O}), Passive (\textit{P}), Active (\textit{A}), Constructive (\textit{C}), or Interactive (\textit{I}). Off-task threads were retained in the statistics of engagement composition for RQ1 but excluded from subsequent analyses. Other on-task threads followed a highest-clearly-evidenced hierarchy: $I > C > A > P$ \citep{chi2014icap}, with higher modes requiring substantive student contribution. Thread-level labels also grouped task sequences for RQ2 and defined comparison groups for RQ3.

\begin{table}[H]
\centering
\caption{Adapted dialogue-oriented ICAP framework.}
\label{tab:icap_framework}
\small

\begin{tabular}{
    p{0.10\textwidth}
    p{0.24\textwidth}
    p{0.29\textwidth}
    p{0.25\textwidth}
}
\toprule
\textbf{Category} &
\textbf{Definition} &
\textbf{Typical indicators} &
\textbf{Coding criteria} \\
\midrule

Off-task &
The interaction is primarily unrelated to academic learning or focuses on platform mechanics. &
Non-academic discussion, interface problems, chatbot testing, prompt injection, or unintelligible input. &
Assigned when the predominant purpose of the complete thread is non-academic or platform-related. \\

Passive &
The student participates in a learning-related interaction but shows little observable cognitive processing. &
Generic requests for help, answers, review, or feedback; acknowledgements without specific academic content. &
Assigned when student messages contain no specific concept, answer, difficulty, attempt, or reasoning. \\

Active &
The student engages with specific academic content without explaining their thinking. &
Specific questions, selected answers, values, error reports, or requests for the next step. &
Assigned when the student provides content-specific information without explanation, justification, critique, or reflection. \\

Constructive &
The student generates reasoning or understanding beyond the information directly provided. &
Explanations, justifications, interpretations, comparisons, critiques, corrections, reflections, or hypotheses. &
Assigned when student messages contain observable student-generated reasoning or conceptual interpretation. \\

Interactive &
The student and AI collaboratively develop understanding across multiple turns. &
Revision after feedback, application of suggestions, iterative clarification, hypothesis testing, or progressive refinement. &
Assigned when sustained student uptake and co-construction are observable across the complete thread. \\

\bottomrule
\end{tabular}
\end{table}

Following evidence that large language models can productively support qualitative codebook construction \citep{barany2024codebook}, ICAP coding followed a five-phase hybrid human-AI workflow: (1) codebook development, (2) coding protocol design, (3) calibration, (4) full corpus coding, and (5) post-hoc human verification. GPT-5.4 drafted an initial codebook, including preliminary category definitions and edge cases, which the research team iteratively revised against conversation logs outside the calibration set until the categories were discriminable in practice. The finalised codebook was then converted into a structured AI prompt incorporating category definitions, hierarchical decision rules, few-shot boundary examples, and an output schema requiring the assigned label, justification, and model-reported confidence. Calibration and full corpus coding used GPT-5.5, released by OpenAI after the codebook was finalised. The coding protocol was calibrated and refined on a sample of 200 threads spanning 1 to 36 turns and all three AI companions. Two researchers and GPT-5.5 coded these threads independently and in a mutually blinded manner. Disagreements were jointly reviewed and the protocol revised over iterative rounds until every ICAP category met pre-specified reliability thresholds (Table~\ref{tab:intercoder_reliability}): pairwise Cohen's $\kappa \geq 0.60$ \citep{cohen1960cefficient} and three-rater Krippendorff's $\alpha \geq 0.70$ \citep{krippendorff2004reliability}, with 10,000-resample bootstrap 95\% confidence intervals (95\% CI). Reliability was highest for Off-task, Constructive, and Interactive and lowest for Active. Disagreements concentrated at the boundary between generic and content-specific requests and at the threshold for student-generated reasoning, and the wider Interactive CI reflects its low prevalence. The fixed protocol then yielded 2,723 classifications across the full corpus, including the calibration threads. Threads with model-reported confidence below 0.9 ($n=262$, 9.6\%) were adjudicated by researchers, and 35 labels were revised, primarily at the Active-Constructive boundary. To assess coding stability, a further 200 AI-coded threads not used in calibration or adjudication were recoded by a calibration researcher blind to the model output. Category-specific agreement (85.5--97.5\%) and $\kappa$ (0.65--0.83) matched calibration levels, indicating no detectable drift.

\begin{table}[H]
\centering
\caption{Final inter-coder reliability on the 200 calibration samples.}
\label{tab:intercoder_reliability}
\scriptsize
\setlength{\tabcolsep}{3.5pt}

\resizebox{\textwidth}{!}{%
\begin{tabular}{lcccccccc}
\toprule
& \multicolumn{3}{c}{\textbf{Category assignments: $n$ (\%)}}
& \multicolumn{3}{c}{\textbf{Pairwise Cohen's $\kappa$ [95\% CI]}}
& \multicolumn{2}{c}{\textbf{Three-rater reliability}} \\
\cmidrule(lr){2-4}
\cmidrule(lr){5-7}
\cmidrule(lr){8-9}

\textbf{Category}
& \textbf{GPT-5.5}
& \textbf{Human 1}
& \textbf{Human 2}
& \textbf{Human 1--Human 2}
& \textbf{GPT-5.5--Human 1}
& \textbf{GPT-5.5--Human 2}
& \textbf{Agreement: $n$ (\%)}
& \textbf{Krippendorff's $\alpha$ [95\% CI]} \\
\midrule

Off-task
& 41 (20.5)
& 32 (16.0)
& 34 (17.0)
& 0.96 $[0.90, 1.00]$
& 0.82 $[0.70, 0.91]$
& 0.85 $[0.75, 0.94]$
& 189 (94.5)
& 0.88 $[0.79, 0.94]$ \\

Passive
& 59 (29.5)
& 68 (34.0)
& 69 (34.5)
& 0.88 $[0.80, 0.94]$
& 0.76 $[0.65, 0.85]$
& 0.70 $[0.59, 0.80]$
& 171 (85.5)
& 0.78 $[0.70, 0.85]$ \\

Active
& 64 (32.0)
& 66 (33.0)
& 61 (30.5)
& 0.83 $[0.74, 0.91]$
& 0.66 $[0.54, 0.76]$
& 0.62 $[0.49, 0.73]$
& 161 (80.5)
& 0.70 $[0.61, 0.78]$ \\

Constructive
& 24 (12.0)
& 24 (12.0)
& 25 (12.5)
& 0.93 $[0.83, 1.00]$
& 0.76 $[0.60, 0.89]$
& 0.79 $[0.64, 0.91]$
& 189 (94.5)
& 0.83 $[0.72, 0.92]$ \\

Interactive
& 12 (6.0)
& 10 (5.0)
& 11 (5.5)
& 0.95 $[0.81, 1.00]$
& 0.71 $[0.44, 0.90]$
& 0.77 $[0.52, 0.94]$
& 194 (97.0)
& 0.81 $[0.62, 0.94]$ \\
\bottomrule
\end{tabular}%
}
\end{table}

\subsubsection{Conversation Statistical and Semantic Analysis (RQ1)}
To characterise patterns of cognitive engagement and how they varied across workflows, we compared ICAP distributions across the three AI companions. Because each companion was tied to one task, differences in composition indicate how engagement varied across task contexts. Of the 2,723 classified conversation threads, 1,894 in which all user messages matched predefined template messages were excluded, because such threads are Passive by rule and reflect use of conversation starters rather than students' own contributions. Template-only sequences were retained in the RQ2 and RQ3 analyses. The RQ1 analytic sample comprised 829 threads (393 Crea, 244 Modi, and 192 Praxie) from 577 students. All five categories, including Off-task, were retained in the statistical denominator. We calculated frequencies and proportions overall and by companion, with 95\% CI from 10,000 student-cluster bootstrap replicates. Pairwise companion differences were expressed as risk differences in percentage points, with confidence intervals obtained using the same bootstrap procedure. Cramér's $V$ \citep{cramer1946mathematical} summarised the magnitude of association between companion and the complete ICAP distribution, both across all three companions and for each companion pair, with 95\% CIs from the same bootstrap procedure. Given the sparse Constructive and Interactive threads, analyses emphasised descriptive distributions, effect sizes, and clustered confidence intervals. ICAP distributions were visualised using a proportionally stacked bar chart.

To complement the categorical analysis for RQ1, we examined whether the content of on-task student messages varied across ICAP categories and AI companions. A thread-level corpus of student-composed messages was analysed using Leximancer \citep{smith2006leximancer}, which derives concepts from term co-occurrence and groups them into themes. All Off-task threads, AI responses, platform-generated starters, HTML wrappers, and non-English messages were removed before aggregation so that concepts reflected students' own contributions. A concept map was displayed at 50\% theme size to balance thematic differentiation with an interpretable overview of the dominant concepts, their thematic groupings, and relationships across the corpus. Documents were tagged by ICAP category and companion, and concept-tag associations were compared using quadrant reports of relative frequency, strength, and prominence, interpreted alongside coded-segment counts to contextualise group-size differences.

\subsubsection{Process Pattern Analysis of AI Use (RQ2)}
To characterise how students incorporated AI into task completion, we built workflow-specific sequential event logs for Creation, Moderation, and Practice by mapping chronologically ordered events onto workflow-specific state spaces (full state definitions in Appendix~\ref{app:state_definitions}). In addition to synthetic start/end states and student-AI chat across all three workflows, each state space captured workflow-specific actions: Creation included drafting and editing content, constructing options, self-evaluation, submission, and AI feedback; Moderation included evaluative feedback, suggestions, comments, and submission; and Practice included correct and incorrect attempts. Consecutive occurrences of the same state were collapsed, except in Practice, where they represented distinct answer attempts. Sequences were grouped by the ICAP label(s) of their embedded conversation threads, and those spanning multiple categories were classified as mixed.

First-order Markov models (FOMMs) \citep{helske2024process} estimated each workflow's transition structure because they quantify which states immediately precede and follow student-AI dialogue and thus represent where dialogue is situated and what follows it. For adjacent current state $i$ and next state $j$, the transition probability is defined as $P_{ij}=P(S_{t+1}=j\mid S_t=i)=\frac{n_{ij}}{\sum_k n_{ik}}$, where $n_{ij}$ denotes the number of observed $i\rightarrow j$ transitions and the denominator sums over all possible next states $k$. These probabilities formed a transition matrix $\mathbf{P}=[P_{ij}]$ for each workflow, in which higher probabilities indicated more prominent immediate pathways and lower probabilities indicated less frequently observed alternatives. Each category-specific matrix was compared with the overall workflow matrix as a baseline to express deviations. For transition $i\rightarrow j$ in category $g$, the probability difference $\Delta P_{ij}^{(g)}=P_{ij}^{(g)}-P_{ij}^{(\mathrm{Overall})}$, where $P_{ij}^{(g)}$ and $P_{ij}^{(\mathrm{Overall})}$ respectively denote the corresponding category-specific and overall transition probabilities (including mixed-category sequences but excluding Off-task sequences), indicated whether the transition was more (positive) or less (negative) prominent in that category than in the overall workflow structure. Comparisons were restricted to the same workflow as each one had a unique state space.

\subsubsection{Outcome Analysis of AI-Supported Learning (RQ3)}
To examine whether dialogue and task-process characteristics corresponded to differences in behavioural re-engagement and task outcomes, each coded thread and its task sequence were linked to submitted multiple-choice artefacts (MCQs) in Creation and individual attempts in Practice, using the same sequences as FOMMs but excluding Off-task and mixed-category sequences, for which no single ICAP category could be assigned. Because students chose when and how to use AI and contributed multiple artefacts and attempts, all models for this analysis estimated covariate-adjusted associations with cluster-robust inference. Full specifications are given in Appendix~\ref{app:model_specs}.

In Creation, outcomes were behavioural re-engagement (count of dialogue-to-task transitions) and artefact quality, rated by peers during Moderation on a five-point scale (\textit{1 = Poor; 2 = Needs improvement; 3 = Satisfactory; 4 = Great; 5 = Outstanding}). The overall decision was the primary quality outcome; correctness and the quality of question, option, and explanation were secondary. Sequences with at least one student-composed message formed the student-composed group (\textit{SC}), which could also contain template messages; sequences with only template messages formed the template-only group (\textit{TO}), coded Passive (\textit{P}) because templates are all generic academic requests according to Table~\ref{tab:icap_framework}. Within \textit{SC}, sparse ICAP categories were combined into Constructive/Interactive (\textit{SC C/I}) and Passive/Active (\textit{SC P/A}), following the theoretical distinction concerning knowledge-generating contributions beyond materials \citep{chi2014icap, chi2018practice}. Two contrasts were examined within course offerings containing both groups: message source (\textit{SC P} versus \textit{TO P}) and engagement depth (\textit{SC C/I} versus \textit{SC P/A}). The two primary links, message source with re-engagement and engagement depth with overall quality, formed a Holm family \citep{holm1979simple}. Each contrast was also estimated on the other outcome as a cross-path check without multiplicity correction. Behavioural re-engagement was modelled with a Gaussian generalised estimating equation (GEE) \citep{liang1986longitudinal, zeger1986longitudinal} clustered by author, yielding adjusted means and mean differences (MD). Artefact quality was modelled with cumulative-logit regression \citep{mccullagh1980regression} on reviewer-level ratings with author $\times$ reviewer cluster-robust covariance, reported as proportional odds ratios (POR). For overall quality, estimates were additionally converted to adjusted expected ratings on the five-point scale. Both models adjusted for authors' prior rating history, submission order, and offering; the quality model also adjusted for reviewer severity and Modi use before rating. Because only final MCQ versions were rated, estimates concern adjusted final quality rather than within-artefact gain. To assess whether the number of dialogue-to-task transitions was itself associated with quality, a process-adjusted model added standardised dialogue-to-task transitions to each quality model, with the other post-dialogue measures entered as conditioning covariates; the transition coefficient was reported as a further cross-path check. The four secondary rubric dimensions for the engagement-depth contrast formed a Benjamini--Hochberg family \citep{benjamini1995controlling}. The proportional-odds assumption was checked with threshold-specific dialogue coefficients, and peer-rating agreement was quantified with linear-weighted Gwet's AC2 \citep{gwet2008reliability, gwet2014handbook}. Robustness was assessed by re-estimating quality on artefact-level mean ratings, excluding ratings preceded by moderator AI use, restricting to artefacts whose authors had at least one previously rated artefact, and omitting each offering in turn.

In Practice, outcomes after an incorrect attempt were continuation (any further attempt), conditional correctness (a correct next attempt among continuers only), and joint correct recovery (a further attempt that was correct among all incorrect attempts). AI exposure (\textit{AI} versus \textit{no AI}) was defined as any dialogue within a 30-second landmark window after the incorrect attempt, with outcomes observed over the following 600 seconds. Incorrect attempts followed by another attempt within the window were excluded so that exposure preceded outcome ascertainment. The three outcomes were estimated by logistic regression with student $\times$ resource two-way cluster-robust covariance, adjusting for learner ability, resource difficulty, mastery rating, attempt position, preceding incorrect attempts, and offering, and were reported as odds ratios (OR) and adjusted probabilities. Robustness was assessed with 15- and 60-second windows and, where estimable, a within-student specification among students observed under both exposure conditions. Because post-error help-seeking is elective, the Practice analysis was exploratory. Its three outcomes decompose a single post-error response and were interpreted jointly without multiplicity correction.

\section{Results}
\subsection{RQ1: Cognitive Engagement With AI}
\subsubsection{ICAP Engagement Patterns Across AI Companions}
ICAP distributions (Table~\ref{tab:icap_distribution}) indicated Passive and Active engagement together accounted for 535 of the 829 conversation threads (64.5\%) and formed the majority of the analytic sample, with Passive the most frequent category (37.4\%) followed by Active (27.1\%). Constructive (14.0\%) and Interactive (13.3\%) engagement were less common, and Off-task interactions were least frequent (8.2\%). Representative complete conversation threads for each category are provided in Appendix~\ref{app:icap_examples}.

\begin{table*}[htbp]
\centering
\caption{Companion-specific ICAP distributions with student-clustered 95\% CI.}
\label{tab:icap_distribution}
\footnotesize
\setlength{\tabcolsep}{4.5pt}
\begin{tabular}{@{}lcccc@{}}
\toprule
\textbf{Category} &
\shortstack{\textbf{Overall}\\($n=829$)} &
\shortstack{\textbf{Crea}\\($n=393$, 47.4\%)} &
\shortstack{\textbf{Modi}\\($n=244$, 29.4\%)} &
\shortstack{\textbf{Praxie}\\($n=192$, 23.2\%)} \\
\midrule

Off-task
& 68 (8.2\%; $[6.2, 10.4]$)
& 27 (6.9\%; $[4.4, 9.6]$)
& 12 (4.9\%; $[2.4, 7.9]$)
& 29 (15.1\%; $[9.4, 21.5]$) \\

Passive
& 310 (37.4\%; $[33.3, 41.6]$)
& 102 (26.0\%; $[21.3, 31.0]$)
& 144 (59.0\%; $[50.8, 67.2]$)
& 64 (33.3\%; $[25.4, 42.2]$) \\

Active
& 225 (27.1\%; $[23.5, 30.8]$)
& 103 (26.2\%; $[21.8, 30.9]$)
& 42 (17.2\%; $[11.3, 23.9]$)
& 80 (41.7\%; $[32.7, 50.8]$) \\

Constructive
& 116 (14.0\%; $[11.3, 16.8]$)
& 76 (19.3\%; $[15.3, 23.5]$)
& 30 (12.3\%; $[7.5, 17.6]$)
& 10 (5.2\%; $[1.1, 10.4]$) \\

Interactive
& 110 (13.3\%; $[10.6, 16.2]$)
& 85 (21.6\%; $[16.9, 26.5]$)
& 16 (6.6\%; $[3.4, 10.1]$)
& 9 (4.7\%; $[2.0, 7.8]$) \\

\bottomrule
\end{tabular}
\end{table*}

\begin{figure}[H]
    \centering
    \includegraphics[width=1\linewidth]{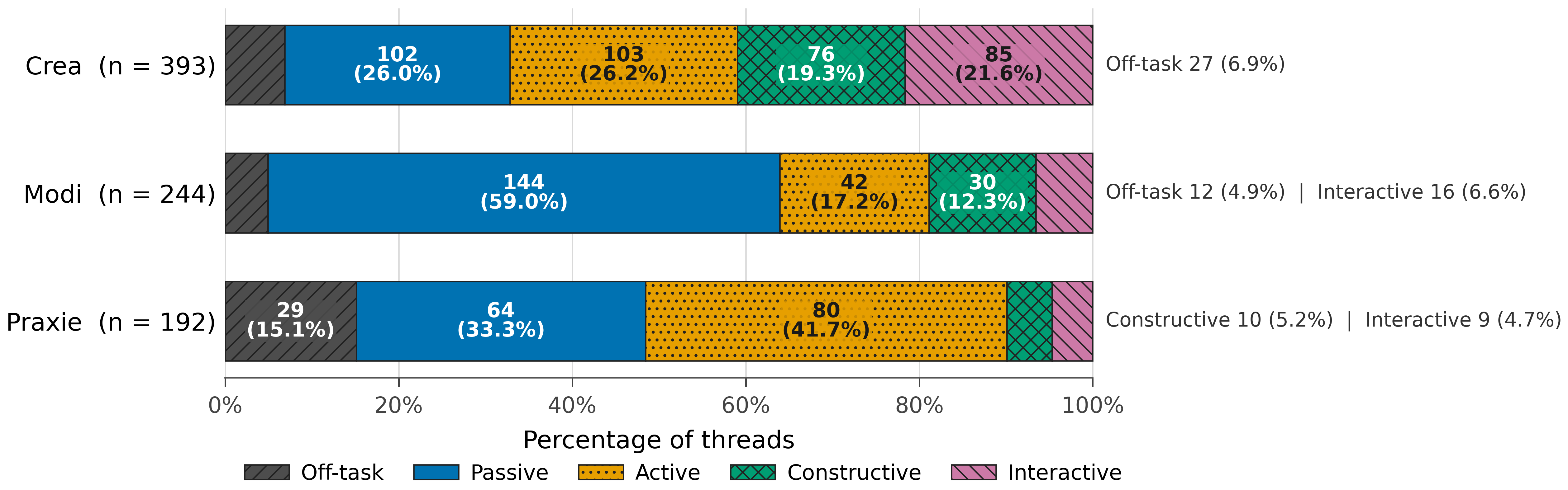}
    \caption{Distribution of ICAP categories by AI companion. Percentages are based on all eligible single-companion threads. Counts and percentages are shown within the bars where space permits; labels for narrow segments are shown on the right.}
    \label{fig:icap_distribution}
\end{figure}

Engagement composition varied across the three AI companions (Figure~\ref{fig:icap_distribution}). Crea showed a comparatively even distribution across four on-task modes and had the largest Constructive (19.3\%) and Interactive (21.6\%) proportions. Modi was dominated by Passive engagement (59.0\%). Praxie had the largest Active (41.7\%) and Off-task (15.1\%) proportions. Student-clustered risk differences quantified these contrasts. Crea had 33 percentage points fewer Passive threads than Modi (95\% CI: $[-42.2, -24.0]$) and 15 points more Interactive threads (95\% CI: $[9.3, 20.8]$). The Modi--Praxie contrast principally involved the lower engagement, with Modi 25.7 points higher on Passive (95\% CI: $[13.2, 37.6]$) and 24.5 points lower on Active (95\% CI: $[-36.0, -13.2]$). The Crea--Praxie contrast instead involved the higher engagement: Crea was 16.9 points higher on Interactive (95\% CI: $[11.4, 22.4]$) and 14.1 points higher on Constructive (95\% CI: $[7.9, 20.2]$), while their Passive proportions did not clearly differ (95\% CI: $[-17.4, 2.1]$). At the overall distributional level, the association between companion and the complete five-category distribution was Cramér's $V=0.294$ (95\% CI: $[0.246, 0.359]$). The largest observed pairwise association was for Modi versus Praxie ($V=0.359$, 95\% CI: $[0.261, 0.479]$), followed by Crea versus Modi ($V=0.343$, 95\% CI: $[0.264, 0.434]$) and Crea versus Praxie ($V=0.326$, 95\% CI: $[0.253, 0.413]$).
 
\subsubsection{Semantic Patterns in On-Task Student-AI Dialogue}
\begin{figure}[H]
    \centering
    \includegraphics[width=0.75\linewidth]{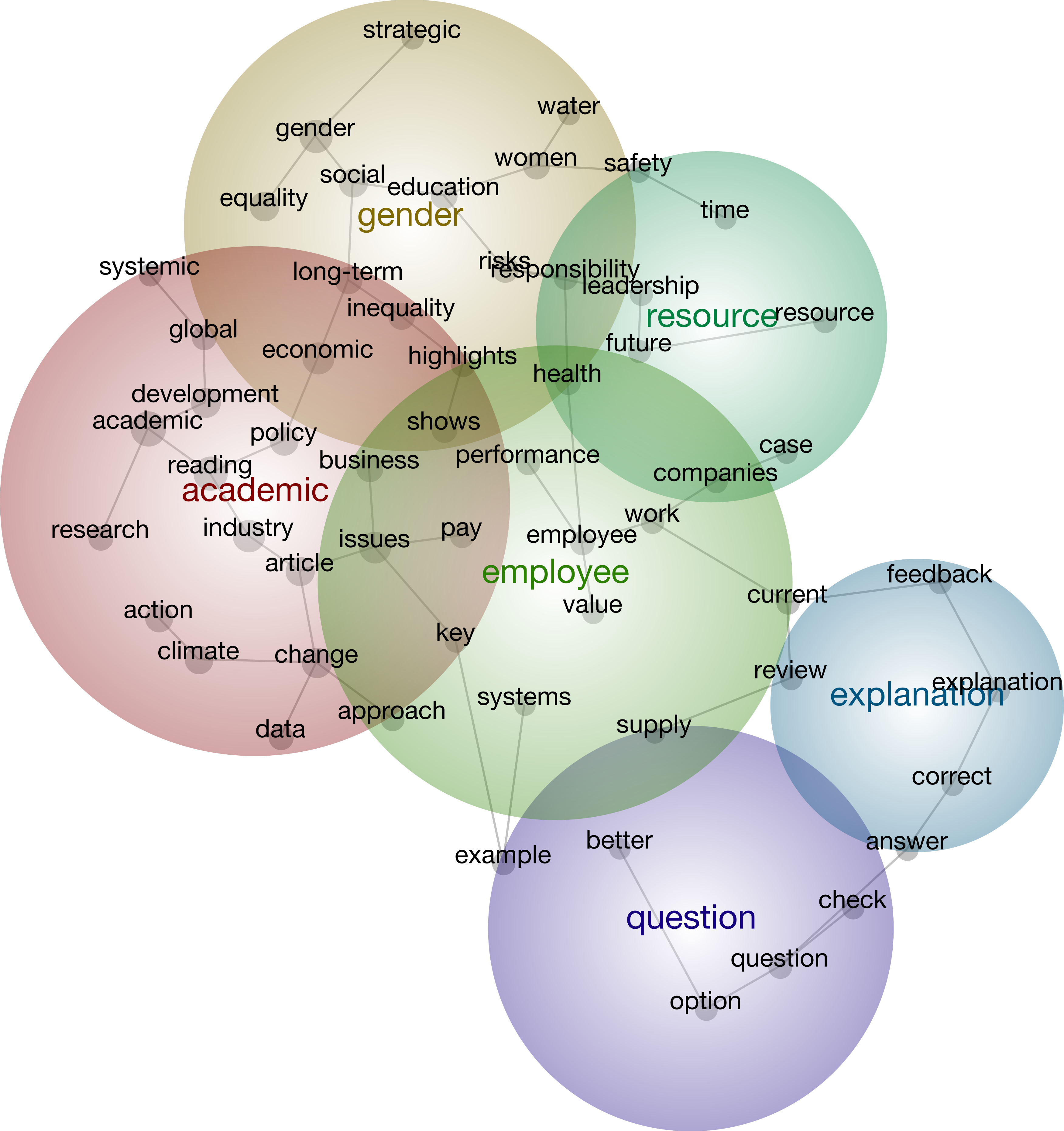}
    \caption{Concept map of on-task student messages (50\% theme size).}
    \label{fig:concept_map}
\end{figure}

The semantic corpus comprised 758 on-task documents, with 750 excerpts retained in the final output (98.9\%; 57 concepts) tagged by companion (360 Crea, 231 Modi, 159 Praxie) and engagement (303 Passive, 224 Active, 115 Constructive, 108 Interactive). The concept map (Figure~\ref{fig:concept_map}) provides a corpus-level overview of six themes spanning disciplinary content (\textit{academic}, \textit{gender}, \textit{employee}), help-seeking behaviour (\textit{question}, \textit{explanation}), and resource development (\textit{resource}).

\begin{figure}[H]
    \centering

    \begin{subfigure}[t]{0.49\textwidth}
        \centering
        \includegraphics[width=\linewidth]{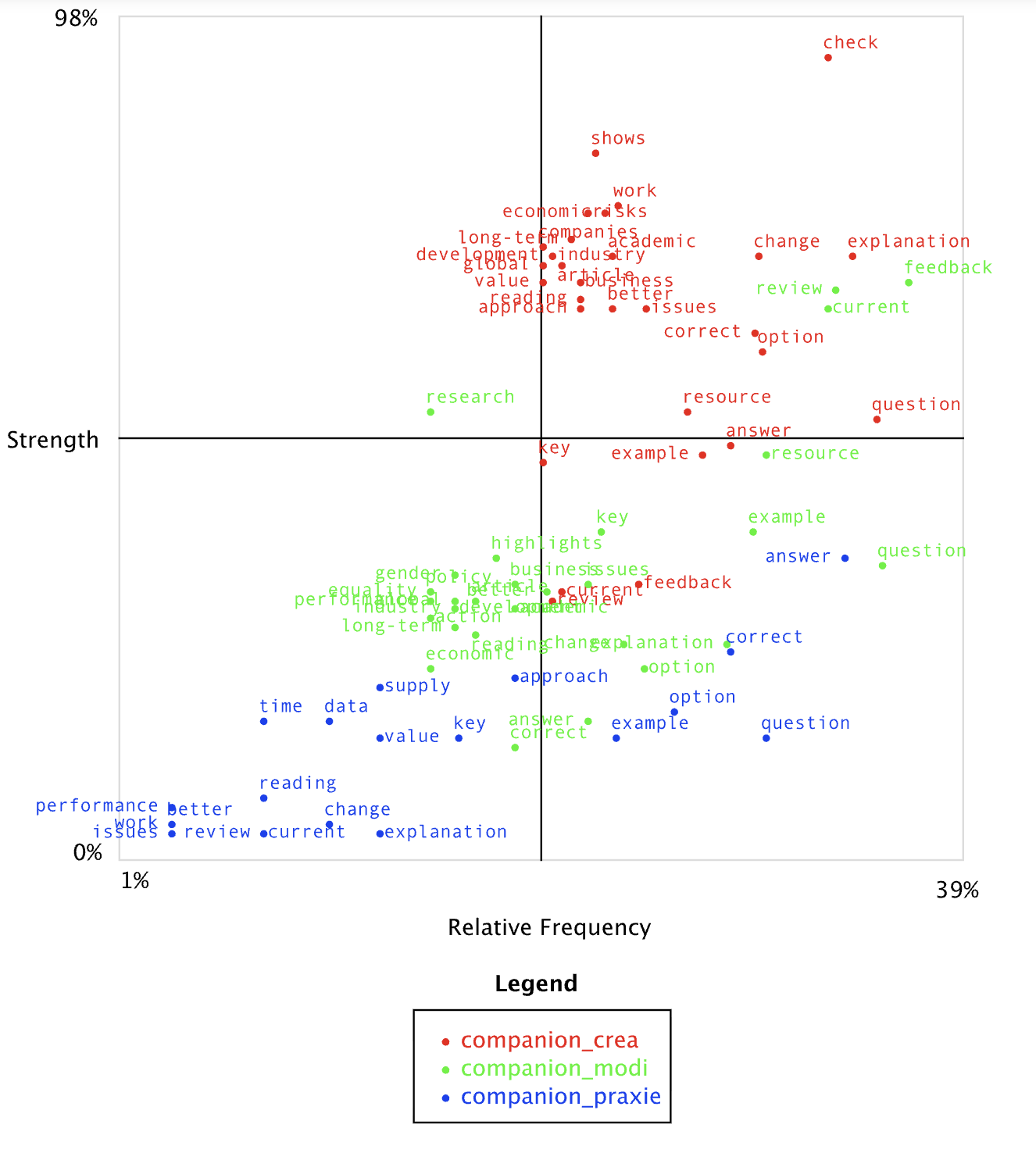}
        \caption{Report by AI companion dimension.}
        \label{fig:companion_quadrant}
    \end{subfigure}
    \hfill
    \begin{subfigure}[t]{0.49\textwidth}
        \centering
        \includegraphics[width=\linewidth]{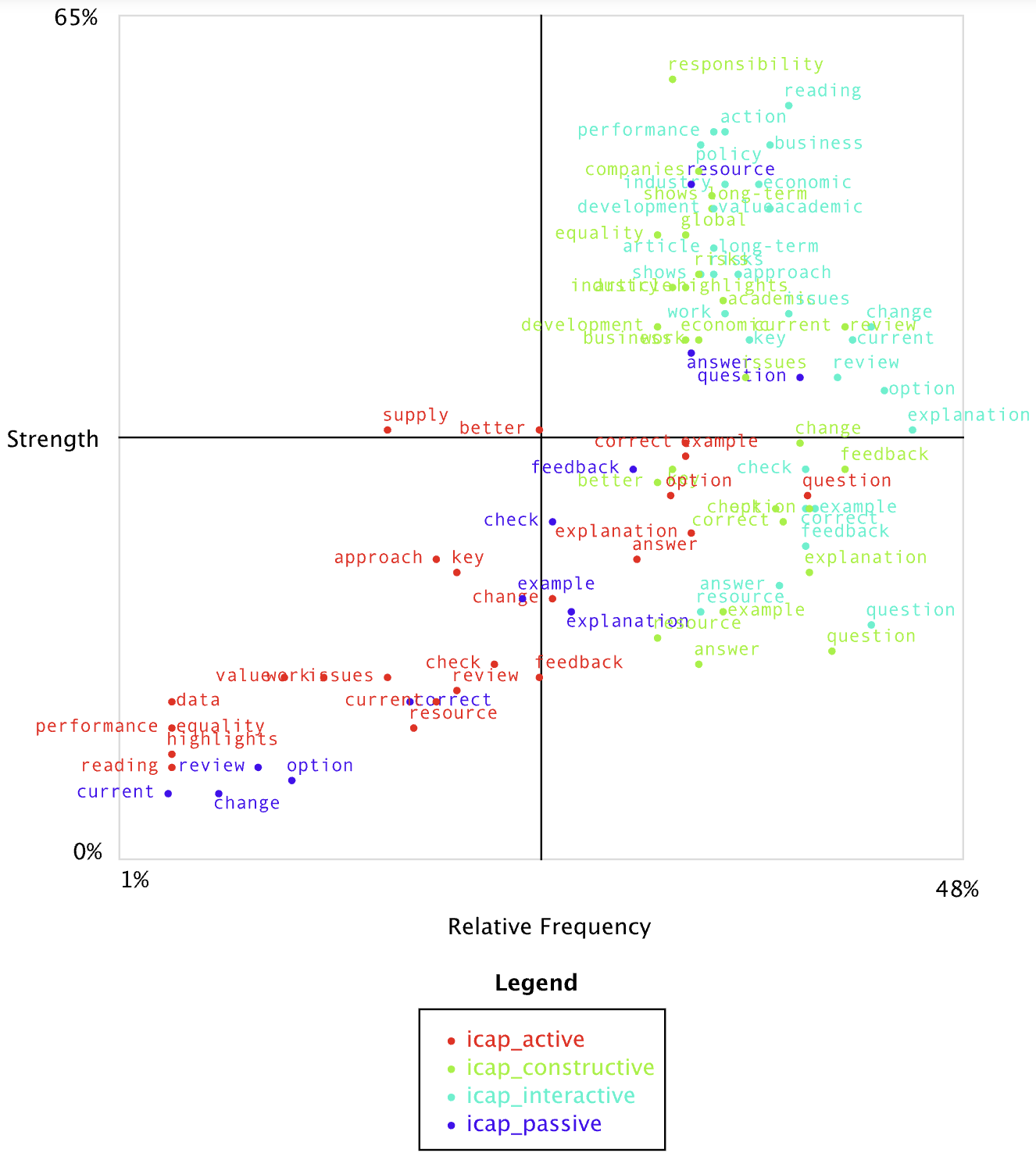}
        \caption{Report by ICAP dimension.}
        \label{fig:icap_quadrant}
    \end{subfigure}

    \caption{Quadrant report for concept strength and relative frequency.}
    \label{fig:quadrant_report}
\end{figure}

Tagged comparisons (Figures~\ref{fig:companion_quadrant}, \ref{fig:icap_quadrant}; Table~\ref{tab:ranked_concepts_for_categories}) differentiated these concepts along both dimensions. By companion, Crea centred on checking and improving drafted resources, Modi on producing evaluative review, and Praxie on obtaining and verifying answers. Across ICAP categories, Passive and Active excerpts showed lower prominence and were dominated by generic help-seeking, task-management, and answer-checking concepts, whereas Constructive and Interactive excerpts showed higher prominence and were distinguished by subject-matter concepts drawn from the specific disciplinary content of the courses. Representative excerpts also illustrated these category-specific patterns: Passive included direct retrieval requests (``what is the correct answer?''); Active, focused clarification requests (``could you explain what 1:N means in this scenario?''); Constructive, tentative reasoning (``if the price is higher, the suppliers want more money, so the supply will increase, right?''); and Interactive, a student challenging an AI-proposed answer by noting that a database rule was inconsistent with a working query and suggesting the question be flagged.

\begin{table}[H]
\centering
\caption{Top-2 ranked concepts. Two highest-prominence concepts for each category are summarised in this table.}
\label{tab:ranked_concepts_for_categories}
\small
\begin{tabular}{lllrrr}
\toprule
\textbf{Dimension} & \textbf{Category} & \textbf{Concept} &
\textbf{Relative frequency (\%)} & \textbf{Strength (\%)} & \textbf{Prominence} \\
\midrule
AI Companion & Crea & \textit{check} & 22 & 93 & 1.9 \\
          &      & \textit{shows} & 8 & 82 & 1.7 \\
          & Modi & \textit{feedback} & 32 & 67 & 2.2 \\
          &      & \textit{review} & 23 & 66 & 2.2 \\
          & Praxie & \textit{answer} & 24 & 35 & 1.7 \\
          &        & \textit{correct} & 14 & 24 & 1.2 \\
\midrule
ICAP & Passive & \textit{resource} & 14 & 52 & 1.3 \\
     &         & \textit{answer} & 14 & 39 & 1.0 \\
     & Active & \textit{correct} & 14 & 32 & 1.1 \\
     &        & \textit{example} & 14 & 31 & 1.1 \\
     & Constructive & \textit{responsibility} & 13 & 60 & 3.9 \\
     &              & \textit{companies} & 15 & 53 & 3.5 \\
     & Interactive & \textit{reading} & 22 & 58 & 4.1 \\
     &             & \textit{performance} & 16 & 56 & 3.9 \\
\bottomrule
\end{tabular}
\end{table}

\subsection{RQ2: Process Patterns of AI Use}
\subsubsection{Analytical Coverage and Overall Transition Structures}
The analysis of process pattern comprised 2,969 task sequences and 13,835 observed state transitions across the three workflows, after excluding 69 sequences classified entirely as Off-task: 1,064 sequences and 5,803 transitions for Creation; 1,104 sequences and 5,099 transitions for Moderation; and 801 sequences and 2,933 transitions for Practice. Table~\ref{tab:fomm_coverage} presents the number of sequences included in the Overall and ICAP-specific models.

\begin{table}[H]
\centering
\caption{Sequence coverage for the Overall and ICAP-specific FOMMs.}
\label{tab:fomm_coverage}
\begin{tabular}{lrrrrrr}
\hline
\textbf{Workflow} &
\textbf{Overall} &
\textbf{Interactive} &
\textbf{Constructive} &
\textbf{Active} &
\textbf{Passive} &
\textbf{Mixed} \\
\hline
Creation   & 1,064 & 114 & 97 & 115 & 737 & 1 \\
Moderation & 1,104 & 29  & 59 & 38  & 973 & 5 \\
Practice   & 801   & 12  & 35 & 151 & 599 & 4 \\
\hline
\end{tabular}
\end{table}

\begin{figure}[H]
    \centering

    \begin{subfigure}{0.37\textwidth}
        \centering
        \includegraphics[width=\linewidth]{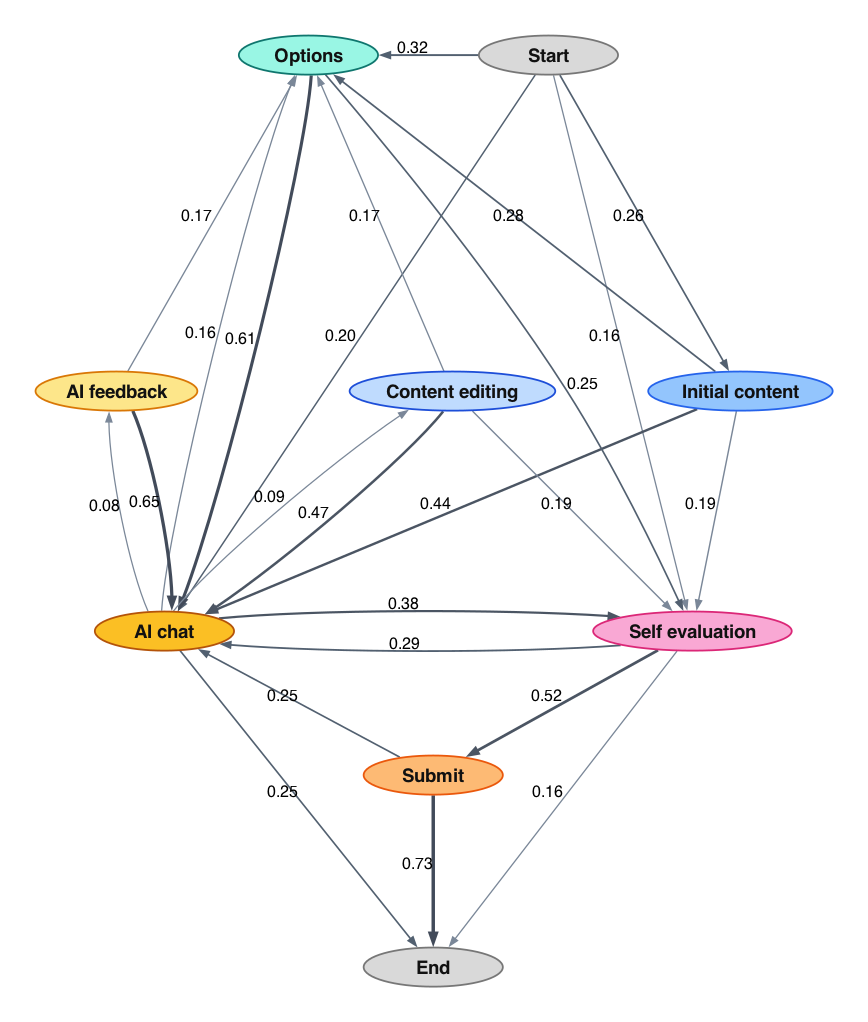}
        \caption{Creation.}
        \label{fig:creation_fomm}
    \end{subfigure}
    \hfill
    \begin{subfigure}{0.34\textwidth}
        \centering
        \includegraphics[width=\linewidth]{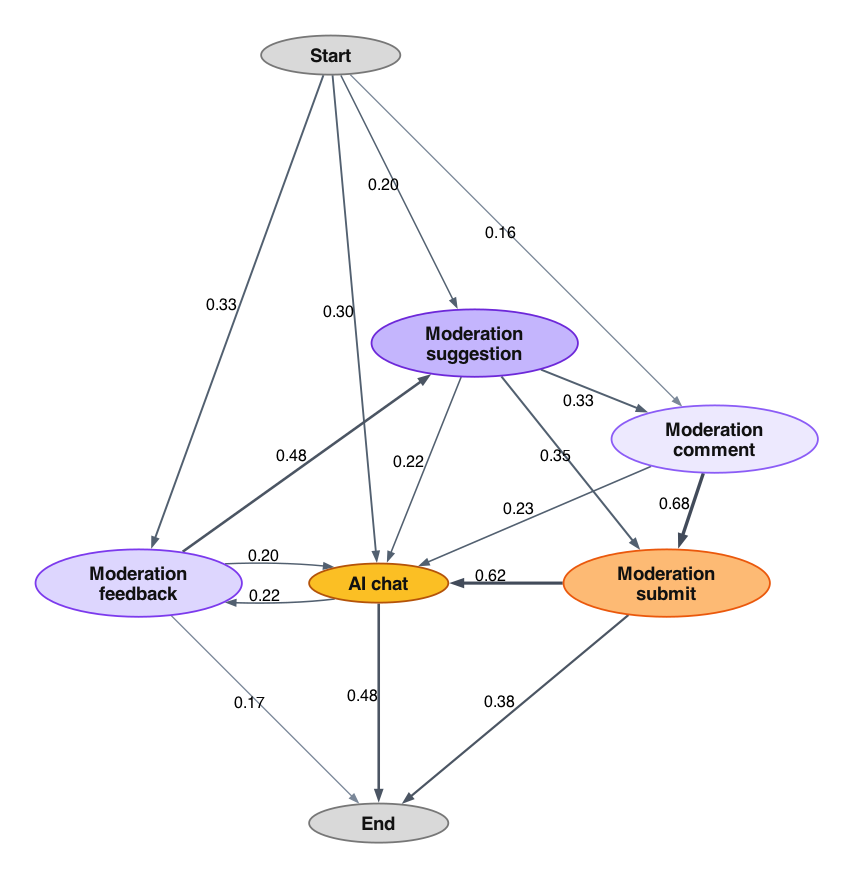}
        \caption{Moderation.}
        \label{fig:moderation_fomm}
    \end{subfigure}
    \hfill
    \begin{subfigure}{0.27\textwidth}
        \centering
        \includegraphics[width=\linewidth]{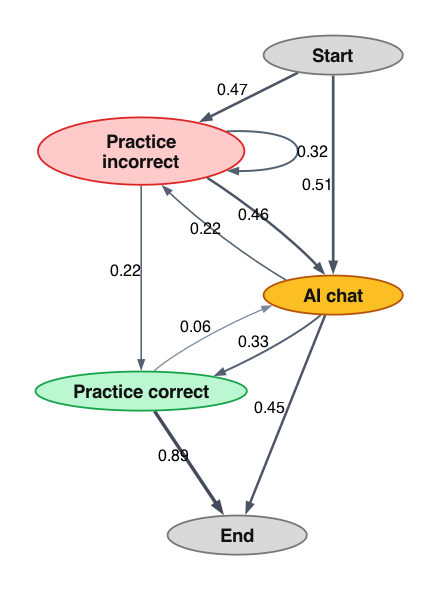}
        \caption{Practice.}
        \label{fig:practice_fomm}
    \end{subfigure}

    \caption{First-order Markov models (FOMMs) for three workflows.}
    \label{fig:overall_fomms}
\end{figure}

Figure~\ref{fig:overall_fomms} presents the Overall first-order Markov model (FOMM) for each workflow, in which nodes denote workflow states and directed edges denote state transitions. Edge labels, widths, and colour intensity indicate the magnitude of transition probability. To improve readability, the diagrams display a filtered subset of the observed transitions while preserving the most salient pathways and preventing isolated intermediate states.

Creation sequences (Figure~\ref{fig:creation_fomm}) most frequently began with option construction and initial drafting from blank fields. Student-AI dialogue occurred throughout the workflow: it could precede authoring and also recur after initial drafting, content editing, option construction, self-evaluation, and automated feedback, with the strongest pattern observed after option construction. Automated feedback led back to dialogue more often than to any authoring state. Following the dialogue, students most frequently moved to self-evaluation or terminated their workflow, with option construction and content revision occurring less often. Self-evaluation was the principal state preceding submission. Submission usually marked the end of the sequence, although it was sometimes followed by renewed dialogue.

Moderation sequences (Figure~\ref{fig:moderation_fomm}) most frequently began with evaluative feedback and a student-AI dialogue before any review content had been produced, while direct suggested improvements or comments were less common entry states. Review activities generally progressed from evaluative feedback to suggested improvements and then towards additional comments or submission. Dialogue remained accessible throughout the workflow, with evaluative feedback, suggested improvements, and additional comments all followed by dialogue in a proportion of cases. Following dialogue, session termination was the most common continuation, while returns to revise the moderation content sometimes occurred. Submission did not necessarily mark the end of the sequence, as it was more frequently followed by re-initiating dialogue.

Compared with Creation and Moderation, Practice sequences (Figure~\ref{fig:practice_fomm}) exhibited relatively less complex transition structures. They usually began with a dialogue before any attempt or an incorrect first attempt. Correct first attempts were uncommon. Incorrect attempts were most frequently followed by dialogue, or another incorrect or correct attempt, while immediate termination was rare. Following dialogue, session termination was the most common next state, with correct attempts occurring more frequently than further incorrect attempts. Once a correct attempt was reached, the sequence usually ended, with additional attempts or renewed dialogue occurring infrequently.

\subsubsection{ICAP-Specific Deviations from the Overall Transition Structures}
\begin{figure}[H]
    \centering
    \includegraphics[width=1\linewidth]{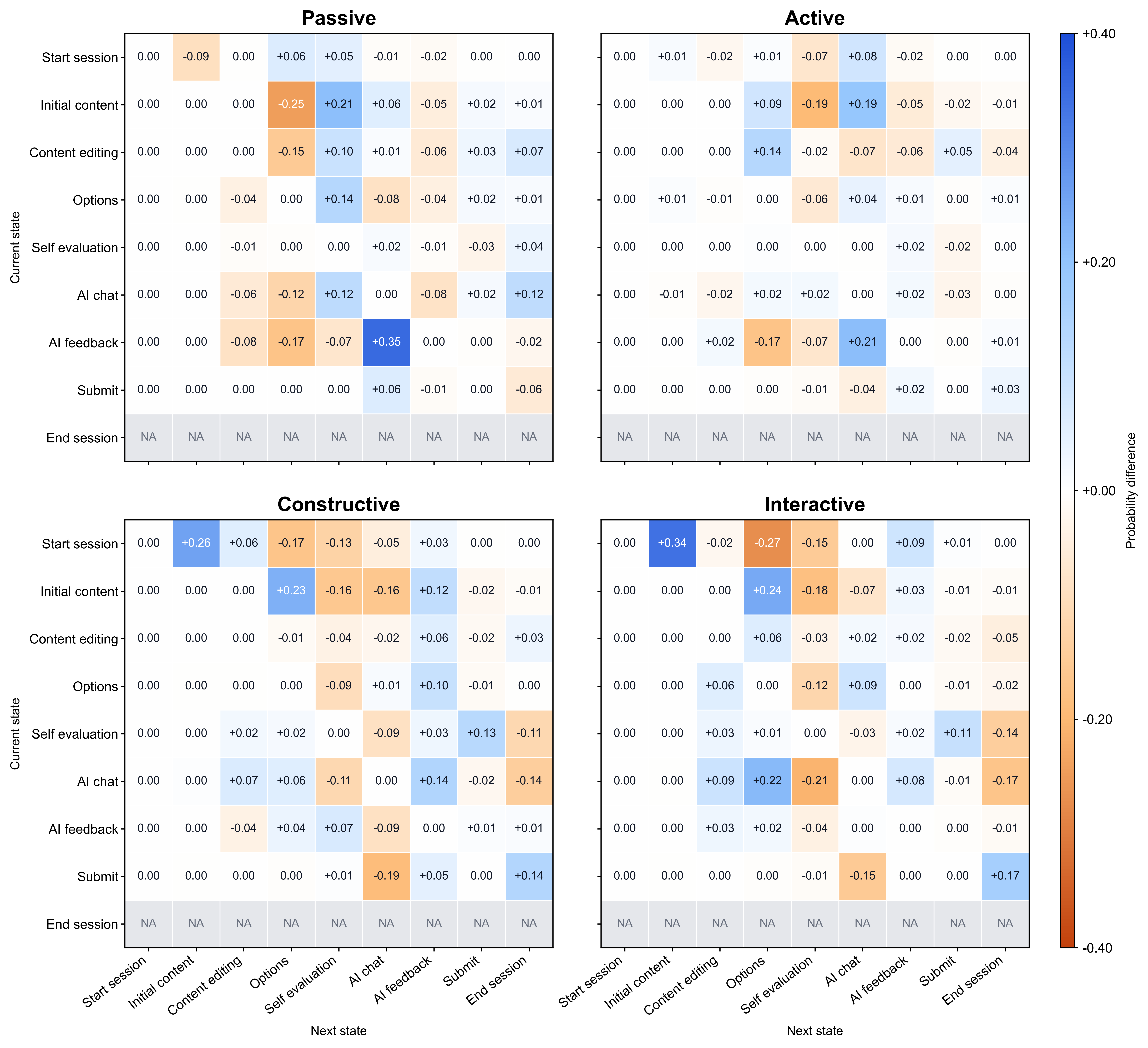}
    \caption{Baseline-adjusted matrices for Creation.}
    \label{fig:creation_matrices}
\end{figure}

In Creation (Figure~\ref{fig:creation_matrices}), Passive sequences moved more often towards self-evaluation, from initial drafting ($+0.21$), option construction ($+0.14$), dialogue ($+0.12$), and content editing ($+0.10$), and less often into option construction, from initial drafting ($-0.25$), automated feedback ($-0.17$), content editing ($-0.15$), and dialogue ($-0.12$). The largest Passive deviation, automated feedback to dialogue ($+0.35$), was based on only three outgoing transitions. Active sequences differed most clearly around initial drafting and automated feedback, both more often followed by dialogue ($+0.19$ and $+0.21$, respectively). Constructive and Interactive sequences departed more in both session entry and post-dialogue activity: both began more frequently with initial drafting ($+0.26$ and $+0.34$, respectively) and less frequently with option construction ($-0.17$ and $-0.27$, respectively). Constructive dialogue was more often followed by automated feedback ($+0.14$) and less often by session termination ($-0.14$), whereas Interactive dialogue more often returned to option construction ($+0.22$) and less often moved to self-evaluation ($-0.21$) or session termination ($-0.17$). After submission, both higher-engagement categories showed less renewed dialogue and more frequent sequence termination.

\begin{figure}[H]
    \centering
    \includegraphics[width=1\linewidth]{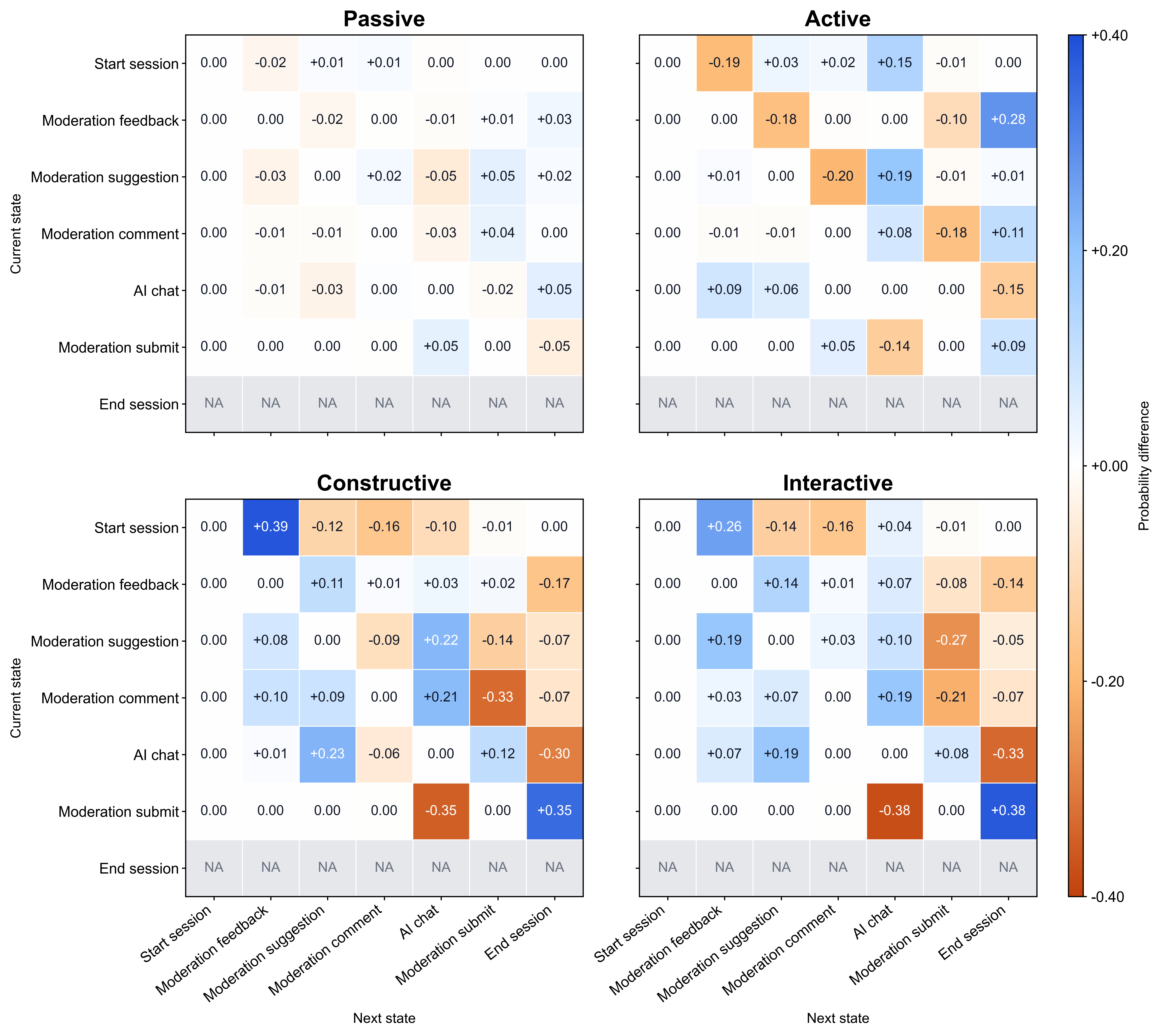}
    \caption{Baseline-adjusted matrices for Moderation.}
    \label{fig:moderation_matrices}
\end{figure}

In Moderation (Figure~\ref{fig:moderation_matrices}), Passive sequences, which constituted the large majority of the workflow (973 of 1,104, 88.1\%), closely tracked the pooled structure. Active sequences departed around session entry and review progression: sessions more often began with dialogue ($+0.15$) and less often with evaluative feedback ($-0.19$). Suggested improvements were more often followed by dialogue ($+0.19$) and less often by additional comments ($-0.20$). Evaluative feedback was more often followed directly by session termination ($+0.28$). Constructive and Interactive sequences showed more pronounced changes in how dialogue was situated within review composition, both starting more often with evaluative feedback ($+0.39$ and $+0.26$, respectively). In Constructive sequences, suggested improvements and additional comments more frequently led to dialogues ($+0.22$ and $+0.21$, respectively), while dialogue more often returned to suggested improvements ($+0.23$) and less often led directly to session termination ($-0.30$). Interactive sequences showed a similar reduction in post-dialogue termination ($-0.33$), with more frequent transitions from suggested improvements to evaluative feedback, from additional comments to dialogue, and from dialogue to suggested improvements ($+0.19$ each). In both higher-engagement categories, submission was substantially more likely to terminate the sequence and less likely to be followed by renewed dialogue, with deviations of approximately $\pm0.35$ for Constructive and $\pm0.38$ for Interactive sequences.

\begin{figure}[H]
    \centering
    \includegraphics[width=1\linewidth]{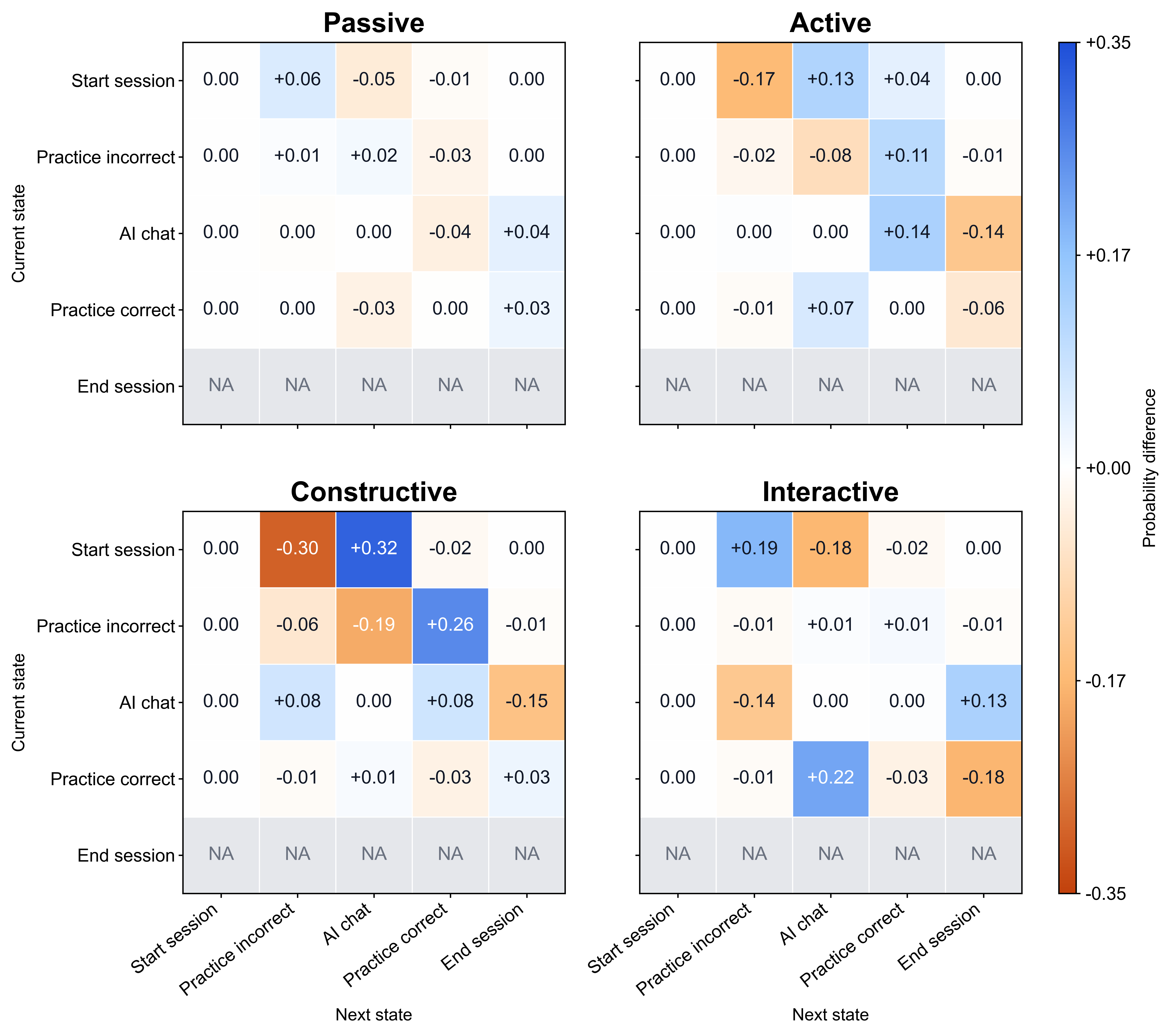}
    \caption{Baseline-adjusted matrices for Practice.}
    \label{fig:practice_matrices}
\end{figure}

In Practice (Figure~\ref{fig:practice_matrices}), Passive sequences, again the majority (599 of 801, 74.8\%), closely tracked the pooled structure. Active sequences more often began with dialogue ($+0.13$) and less often with an incorrect attempt ($-0.17$). Following dialogue, correct attempts were more prominent ($+0.14$) and session termination was less prominent ($-0.14$). Incorrect attempts were more often followed by a correct attempt ($+0.11$). Constructive sequences showed larger deviations in the same directions: sessions much more often began with dialogue ($+0.32$) and much less often with an incorrect attempt ($-0.30$), incorrect attempts more frequently transitioned to correct ones ($+0.26$) and less frequently to dialogue ($-0.19$), and dialogue was less often followed by session termination ($-0.15$). The Interactive model, estimated from 12 sequences with sparse transition rows, is reported descriptively: it showed more frequent session starts with an incorrect attempt ($+0.19$) and more frequent dialogue following a correct attempt ($+0.22$).

\subsection{RQ3: Post-Dialogue Activity and Task Outcomes}
\subsubsection{Creation Outcomes: Dialogue, Process, and Artefact Quality}
\begin{figure}[H]
    \centering
    \includegraphics[width=1\textwidth]{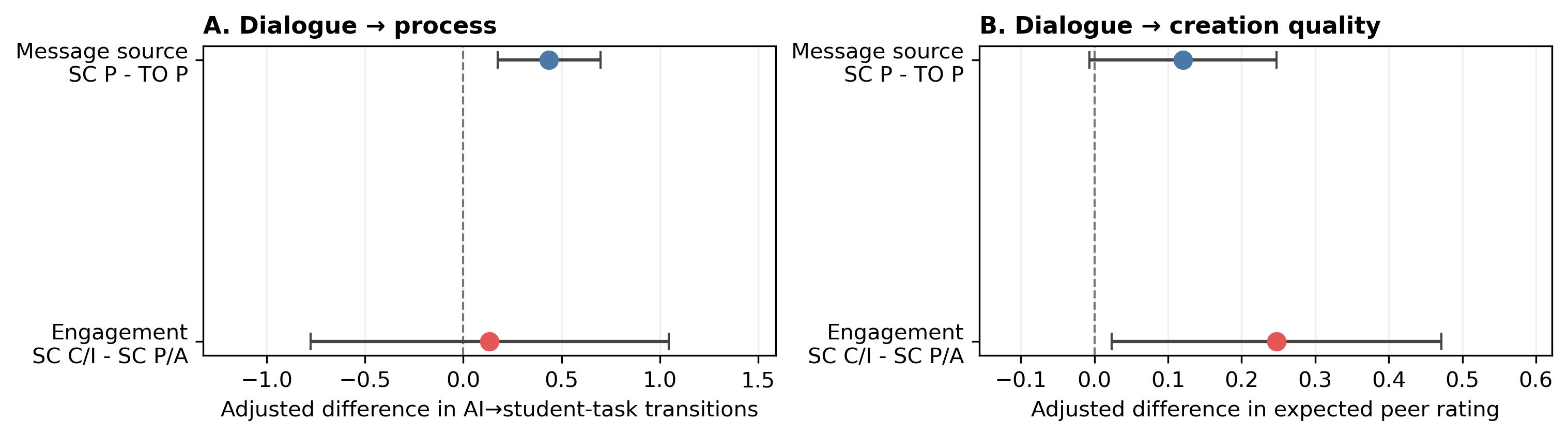}
    \caption{Adjusted dialogue contrasts on the Creation process and quality outcomes (95\% CI). Numerical estimates are reported in Table~\ref{tab:outcome_analysis}.}
    \label{fig:creation_outcome}
\end{figure}

The analysis of resource creation started from the 1,064 Creation sequences reported in Table~\ref{tab:fomm_coverage} and retained 346 submitted MCQs that satisfied the linkage and peer-rating requirements, authored by 281 students and carrying 1,133 overall ratings. The message-source contrast covered 192 artefacts (167 \textit{TO P}, 25 \textit{SC P}) and the engagement contrast covered 98 artefacts (34 \textit{SC P/A}, 64 \textit{SC C/I}). Table~\ref{tab:outcome_analysis} reports all estimates, and Figure~\ref{fig:creation_outcome} places the two dialogue contrasts side by side on the process and quality scales.

\begin{table}[H]
\centering
\caption{Dialogue-process-outcome estimates for Creation and landmark outcomes for Practice.}
\label{tab:outcome_analysis}
\begin{threeparttable}

\small
\setlength{\tabcolsep}{4pt}
\renewcommand{\arraystretch}{1.08}

\begin{tabularx}{\textwidth}{
    >{\raggedright\arraybackslash}p{3.1cm}
    >{\raggedright\arraybackslash}X
    >{\centering\arraybackslash}p{1.15cm}
    >{\centering\arraybackslash}p{1.15cm}
    >{\centering\arraybackslash}p{1.75cm}
    >{\centering\arraybackslash}p{1.45cm}
}
\toprule
\textbf{Contrast}
& \textbf{Outcome}
& \textbf{Effect}
& \textbf{Estimate}
& \textbf{95\% CI}
& \textbf{$\boldsymbol{p}$ ($\boldsymbol{p}_{\mathrm{adj}}$)} \\
\midrule

\multicolumn{6}{l}{\textbf{\textit{Panel A. Creation}}} \\
\addlinespace[2pt]

\textit{SC P} vs \textit{TO P}\textsuperscript{*}
& Dialogue-to-task transitions
& MD
& 0.44
& $[0.18, 0.70]$
& .001 (.002) \\

\textit{SC P} vs \textit{TO P}\textsuperscript{\dag}
& Overall rating
& POR
& 1.46
& $[0.99, 2.17]$
& .059 \\

\textit{SC C/I} vs \textit{SC P/A}\textsuperscript{\dag}
& Dialogue-to-task transitions
& MD
& 0.13
& $[-0.78, 1.04]$
& .774 \\

\textit{SC C/I} vs \textit{SC P/A}\textsuperscript{*}
& Overall rating
& POR
& 2.12
& $[1.08, 4.16]$
& .029 (.029) \\

Transitions (per SD)\textsuperscript{\dag}
& Overall rating, message-source comparison sample
& POR
& 0.91
& $[0.58, 1.43]$
& .697 \\

Transitions (per SD)\textsuperscript{\dag}
& Overall rating, engagement comparison sample
& POR
& 0.78
& $[0.25, 2.46]$
& .673 \\

\midrule
\addlinespace[2pt]
\multicolumn{6}{l}{\textbf{\textit{Panel B. Practice}}} \\
\addlinespace[2pt]

\textit{AI} vs \textit{no AI} in landmark\textsuperscript{\ddag}
& Continuation
& OR
& 0.37
& $[0.16, 0.83]$
& .017 \\

\textit{AI} vs \textit{no AI} in landmark\textsuperscript{\ddag}
& Correctness given continuation
& OR
& 2.78
& $[1.17, 6.63]$
& .021 \\

\textit{AI} vs \textit{no AI} in landmark\textsuperscript{\ddag}
& Joint correct recovery
& OR
& 1.72
& $[0.81, 3.65]$
& .156 \\

\bottomrule
\end{tabularx}

\begin{tablenotes}[flushleft]
\footnotesize
\item \textit{Note}.
\textit{SC} = sequences containing at least one student-composed message; \textit{TO} = sequences containing only predefined template messages.
MD = mean difference; POR = proportional odds ratio for a higher peer rating; OR = odds ratio for the stated binary outcome.
\textsuperscript{*}Primary;
\textsuperscript{\dag}cross-path;
\textsuperscript{\ddag}exploratory.
\textit{Panel~A} covers 192 artefacts on the message-source contrast and 98 on the engagement contrast;
\textit{Panel~B} covers 259 landmark-eligible incorrect attempts.
Adjusted $p$-values apply the Holm correction across the two primary links.
The four secondary rubric dimensions form a separate Benjamini--Hochberg family
and are reported in the text.
No multiplicity correction was applied within \textit{Panel~B}.
\end{tablenotes}

\end{threeparttable}
\end{table}

The two contrasts behaved differently on the two outcomes. On the re-engagement outcome, \textit{SC P} artefacts recorded 0.44 more dialogue-to-task transitions than \textit{TO P} artefacts (adjusted means 1.39 and 0.95 per artefact; Holm-adjusted $p=.002$), whereas the engagement contrast showed no clear difference (adjusted means 2.18 and 2.05). On the quality outcome, the pattern reversed: \textit{SC C/I} artefacts were more likely than \textit{SC P/A} artefacts to receive a higher rating (POR $=2.12$; Holm-adjusted $p=.029$), corresponding to adjusted expected ratings of 4.33 and 4.08, a difference of 0.25 rating points (95\% CI: $[0.02, 0.47]$), whereas the message-source difference was smaller (0.12 rating points, 95\% CI: $[-0.01, 0.25]$). When process measures were added to the quality models, the number of dialogue-to-task transitions showed no clear association with rating in either comparison sample (PORs 0.91 and 0.78 per standard deviation of dialogue-to-task transitions).

The engagement estimate was consistent in direction and magnitude across the estimable sensitivity analyses. Peer moderators showed substantial agreement on overall decision (AC2 $=0.71$, 95\% CI: $[0.69, 0.74]$) and on the four secondary dimensions (AC2 $=0.68$--$0.72$), and the proportional-odds diagnostic was not rejected for either contrast (message-source $p=.604$; engagement $p=.493$). Omitting each contributing offering in turn returned PORs between 1.66 and 2.38, excluding ratings submitted after moderator AI use returned POR $=2.12$ (95\% CI: $[1.08, 4.15]$), and artefact-level mean ratings yielded a difference of similar magnitude (0.22 rating points, 95\% CI: $[-0.01, 0.45]$). The restriction to authors with at least one previously rated artefact was estimable for the message-source contrast only, returning POR $=1.90$ (95\% CI: $[1.00, 3.64]$). All four secondary rubric dimensions showed PORs above unity, from 1.53 for option quality to 1.77 for correctness, without reaching significance after Benjamini--Hochberg correction (minimum adjusted $p=.192$).

\subsubsection{Practice Outcomes After Incorrect Attempts}
\begin{figure}[H]
    \centering
    \includegraphics[width=1\textwidth]{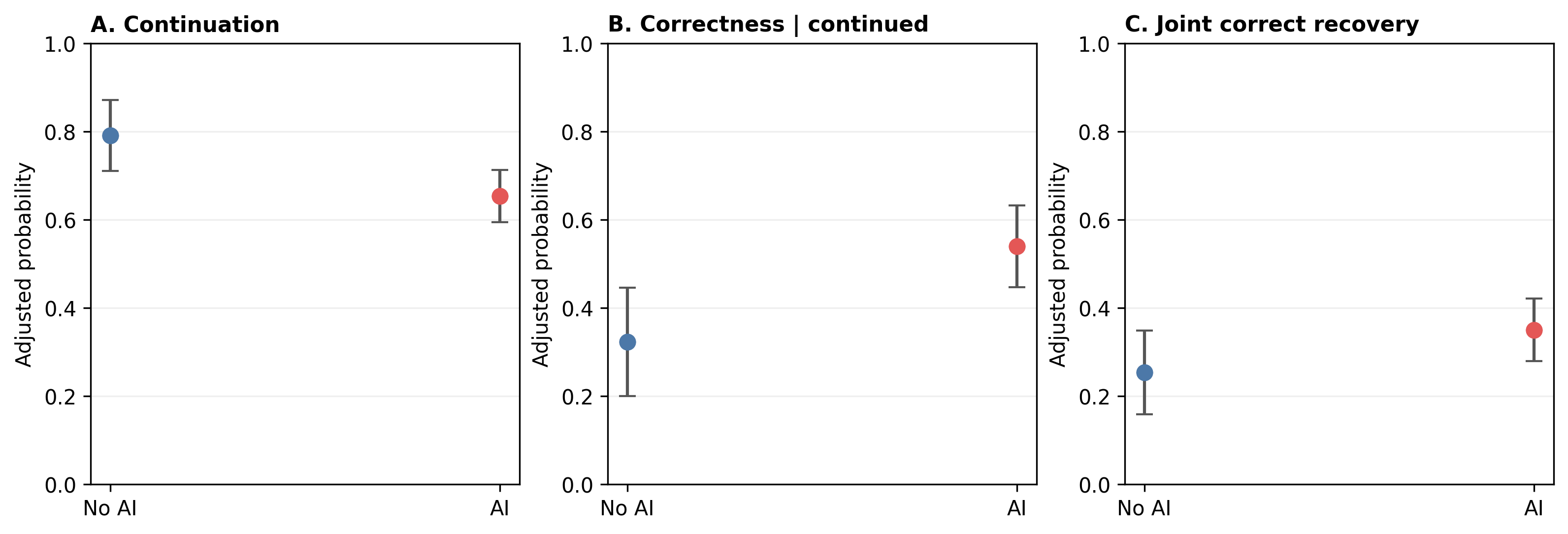}
    \caption{Adjusted probabilities of the three Practice outcomes at the 30-second landmark after an incorrect attempt (95\% CI). Odds ratios for the same comparisons are reported in Table~\ref{tab:outcome_analysis}.}
    \label{fig:practice_outcome}
\end{figure}

The Practice analysis started from the 801 Practice sequences reported in Table~\ref{tab:fomm_coverage} and yielded 259 landmark-eligible incorrect attempts contributed by 166 students, with AI dialogue occurring within the 30-second landmark window for 188 opportunities and absent for 71. The three estimates in \textit{Panel~B} of Table~\ref{tab:outcome_analysis} did not point in a common direction: AI dialogue was associated with lower continuation (OR $=0.37$) but higher conditional correctness (OR $=2.78$), both intervals excluding the null, whereas joint correct recovery was more uncertain (OR $=1.72$, interval covering the null). On the probability scale (Figure~\ref{fig:practice_outcome}), the reduction in continuation and the increase in conditional correctness are of comparable magnitude and their intervals do not overlap, while the two joint-recovery intervals overlap substantially. All three directions were preserved at the 15-second landmark. At 60 seconds, continuation returned OR $=0.34$ (95\% CI: $[0.13, 0.89]$) and joint correct recovery OR $=3.45$ (95\% CI: $[0.89, 13.34]$), while the conditional-correctness model could not be estimated stably because of the reduced number of continuing opportunities. Only 34 students contributed opportunities under both exposure conditions. Within this subgroup, the continuation estimate was imprecise (OR $=0.46$, 95\% CI: $[0.06, 3.48]$) and the remaining two within-person models could not be estimated stably.

\section{Discussion}
Research on student-AI interaction has typically analysed cognitive engagement in dialogue \citep{weatherholtz2025cognitive} and sequences of GenAI use \citep{chen2025unpacking} as separate analytic perspectives. More broadly, studies of student-AI interaction continue to rely substantially on aggregate indicators such as interaction frequency \citep{alamray2026explore, debets2025review, misiejuk2025mapping}. By linking these perspectives in task-embedded AI, this study shows that dialogue characteristics, task-process patterns, and task outcomes capture different aspects of AI use. Cognitive engagement varied across task contexts and was associated with distinct post-dialogue transitions. Message source was associated with behavioural re-engagement, and deeper cognitive engagement with higher final artefact quality. In Practice, AI dialogue after an incorrect attempt was associated with lower continuation but higher correctness among students who continued. Measures of AI use intensity would conflate these distinctions. The following sections examine this in relation to prior work and interpret these findings in relation to the three research questions.

\subsection{Cognitive Engagement Across Task Contexts}
Cognitive engagement with task-embedded AI varied systematically across workflows. Passive and Active modes predominated overall, as in classroom observations \citep{vosniadou2023icap_cg} and GenAI tutoring \citep{weatherholtz2025cognitive}, but their share depended on the task: resource creation involved more Constructive and Interactive dialogue, peer moderation more Passive dialogue, and practice more Active dialogue. Each task plausibly afforded different forms of contribution. In Creation, students could bring their own draft to the exchange, which feedback could prompt them to explain, defend, and revise across turns. In Moderation, the object was a peer's resource, and students could ask the companion to supply evaluative judgement rather than support their own, consistent with reliance on AI assistance in peer feedback \citep{darvishi2024impact}. In Practice, questions with a single correct answer invited content-specific clarification and answer checking. ICAP theory holds that engagement modes are elicited by the design of learning activities \citep{chi2018practice}; in task-embedded AI, the companion's role and the task it supports are part of that design. Because each companion was tied to one task, the differences reflect task demands, companion role, and platform affordances jointly and do not indicate differences in companion effectiveness.

Differences across ICAP modes were not merely distributional. In the semantic analysis, Passive and Active excerpts were more strongly associated with generic help-seeking, task-management, and answer-checking concepts, whereas Constructive and Interactive excerpts were distinguished by more prominent disciplinary concepts drawn from the courses. Thus, higher engagement was distinguished not simply by the presence of academic content, but by more prominent disciplinary concepts and student-generated reasoning around that content. Prior ICAP-based analyses of GenAI dialogue have largely examined a single domain or tutoring format \citep{weatherholtz2025cognitive}, while elaborate AI responses can create an appearance of engagement that does not necessarily reflect students' own cognitive contribution \citep{alamray2026explore}. By analysing students' own contributions across multiple disciplines, the present study shows that ICAP modes remain distinguishable in authentic task-embedded student-AI dialogue while also varying systematically with the context in which that dialogue occurs. This contextual dependence matters for interpretation: a Passive-dominated profile may reflect what a particular task and companion invite instead of students' general capacity for deeper engagement, and similar amounts of AI dialogue can involve different forms of cognitive engagement across workflows.

\subsection{Process Patterns of AI Use Across Task Workflows}
The role of student-AI dialogue depended on where it occurred in the task process. Dialogue recurred across authoring and often followed automated feedback in Creation, accompanied the composition of review content in Moderation, and either preceded a first attempt or followed an incorrect one in Practice. Thus, the same act of consulting AI may take on different functions depending on its task position, such as support before authoring or attempting a problem, consultation around feedback, or help following an error. Help-seeking research distinguishes help that supports students' own problem solving from help that substitutes for it \citep{aleven2003help_seeking}, and whether dialogue precedes or follows a student's own attempt bears on this distinction. More broadly, temporal learning analytics holds that an action's meaning depends on its position in a sequence \citep{knight2017temporal}, which aggregate interaction measures disregard by treating every AI interaction as equivalent. Because workflow design, companion affordances, and students' choices co-occur in this setting \citep{cukurova2025interplay}, these positions describe how dialogue was used in each workflow, not the effect of any single factor. Nonetheless, they indicate that analyses of task-embedded AI use need to record where dialogue occurs, not only how often \citep{chen2025unpacking}.

Within each workflow, cognitive engagement was also associated with different post-dialogue transition patterns. Constructive and Interactive dialogue tended to be less often followed immediately by session termination, but the task activity associated with continued engagement differed by workflow: for example, automated feedback or option construction in Creation, suggested improvements in Moderation, and a further attempt in Practice. Therefore, the important point is not that higher engagement corresponded to a single common pathway, but that engagement within dialogue was associated with different ways of continuing the surrounding task. ICAP-based analyses of student-AI interaction have classified engagement within conversations \citep{weatherholtz2025cognitive}, whereas process analyses of GenAI use have typically treated AI consultation as a common process state \citep{fan2025laziness, chen2025unpacking}. Linking the two shows that dialogues classified at different levels of cognitive engagement were embedded in different immediate post-dialogue task transitions, and that the form of these differences depended on the workflow. Because FOMMs capture immediate transitions \citep{helske2024process}, these associations are descriptive and partly reflect more invested students both engaging more deeply and continuing the task. Apart from how much, the differences also concern which activity followed dialogue, which is why transition patterns and artefact-level return frequency provide complementary information.

\subsection{Dialogue and Process Characteristics in Relation to Task Outcomes}
\subsubsection{Message Source and Behavioural Re-Engagement}
Within Creation sequences classified as Passive, those containing student-composed messages were associated with more behavioural re-engagement, measured as dialogue-to-task returns, than template-only sequences. Both groups shared the Passive label and differed in message source: whether students formulated any part of the request themselves. One plausible explanation is that formulating a request, a step in help-seeking that templates bypass, requires students to initiate and frame the request themselves at that point in the task \citep{aleven2003help_seeking}, which may keep the dialogue tied to the work they return to. The finding extends concerns that conversation starters and prompt suggestions can inflate interaction volume without reflecting student initiative \citep{alamray2026explore, lee2025prompt, park2024promise_peril_chatgpt_he}: within the same ICAP category, whether students formulated their own requests carried information about subsequent task activity that the ICAP label did not.

\subsubsection{Cognitive Engagement, Return Frequency, and Artefact Quality}
Among sequences containing student-composed messages, artefacts associated with Constructive or Interactive dialogue received modestly higher peer ratings of final quality than those associated with Passive or Active dialogue. Because all compared sequences contained student-composed messages, this contrast concerns what students contributed within dialogue. ICAP theory proposes that Constructive and Interactive engagement involve qualitatively deeper cognitive processing than Passive and Active engagement \citep{chi2014icap, chi2018practice}, an ordering recently linked to learning outcomes in GenAI tutoring \citep{weatherholtz2025cognitive}. The present finding extends this ordering to the quality of work students produce with AI during authentic tasks, and it also sharpens evidence that the quality of students' work depends on how students interact with AI rather than on use alone \citep{bai2025impact_genai}: the higher-rated group was characterised by student-generated explaining, justifying, critiquing, or taking up AI input. Feedback literacy research identifies such evaluation and uptake as conditions for feedback to be used \citep{carless2018feedback}, in contrast to the limited scrutiny with which students may adopt AI suggestions \citep{fan2025laziness, zhai2024effects}.

The association between cognitive engagement and artefact quality was not accounted for by the frequency of dialogue-to-task returns. Constructive/Interactive and Passive/Active artefacts did not differ clearly in the number of dialogue-to-task returns, and return frequency was not clearly associated with quality. This pattern is consistent with ICAP's premise that the mode of engagement, rather than the amount of overt activity, distinguishes outcomes \citep{chi2014icap}, and with evidence that repeated AI use can reflect either iterative reasoning or repeated answer-seeking \citep{bai2025impact_genai, zhai2024effects}. A count also collapses what follows dialogue. The transition models showed that Constructive dialogue more often led to automated feedback and Interactive dialogue to option construction. Beyond these transitions, a single return could involve evaluating feedback, critically modifying AI input, or making a superficial change, examples of distinctions that a count cannot register \citep{pozdniakov2025cocreation}. Therefore, return frequency alone cannot represent the form of students' post-dialogue activity, and it did not account for the association between deeper cognitive engagement and final artefact quality in this study.

\subsubsection{Continuation and Conditional Correctness in Practice}
After an incorrect practice attempt, AI dialogue was associated with lower continuation but higher next-attempt correctness among students who continued. Two explanations are consistent with this divergence. Dialogue may have resolved the immediate difficulty for some students, making a further attempt unnecessary, while supporting a correct next attempt for others. Alternatively, because fewer students continued after dialogue, those who continued could have constituted a selected group, which would raise conditional correctness independently of any benefit from the dialogue itself. Joint correct recovery, which does not condition on continuation, therefore prevents higher conditional correctness from being interpreted as evidence of better overall recovery after AI dialogue. In either case, higher next-attempt correctness reflects performance during AI-supported task completion, which does not necessarily indicate learning that persists once support is withdrawn \citep{darvishi2024impact, fan2025laziness, yan2025gains}. This distinction is important when interpreting post-dialogue outcomes in practice: continuation and conditional correctness should not be treated as interchangeable indicators of AI help-seeking. Lower continuation after dialogue need not signal disengagement or unsuccessful support, while higher correctness among students who continued cannot be generalised to all students who sought AI assistance.

\subsection{Distinguishing Post-Dialogue Activity From Task Outcomes}
Across Creation and Practice, dialogue characteristics were associated with post-dialogue activity and task outcomes in different ways. In Creation, message source was associated with behavioural re-engagement and less clearly with final artefact quality, whereas deeper cognitive engagement was associated with higher final artefact quality but not with more dialogue-to-task returns. The number of such returns was not itself associated with quality. In Practice, AI dialogue after an incorrect attempt was associated with lower continuation but higher correctness among students who continued. Therefore, post-dialogue activity and task outcomes are not interchangeable indicators: how often students returned to or continued a task did not indicate how well they performed. Accordingly, the two analytical perspectives were complementary: ICAP coding identified the form of cognitive engagement evident in students' contributions, whereas process analysis identified where dialogue occurred and what task activity followed it. Transition counts alone would not have shown that deeper engagement corresponded to higher artefact quality, and because both message-source groups were Passive, ICAP labels alone would not have distinguished the sequences followed by more re-engagement.

\section{Implications}
The findings of this study highlight the need to treat post-dialogue activity and task outcomes as distinct aspects of AI-supported learning. Accordingly, task-embedded AI support should be designed and evaluated with explicit attention to each, without assuming that improving one improves the other. Prompt suggestions lower the effort required to seek assistance \citep{park2024promise_peril_chatgpt_he}, and AI system designs should also preserve opportunities for students to exercise agency in articulating their own needs \citep{alfredo2024hcla_aied_lr}. Passive dialogue containing student-composed messages was followed by more dialogue-to-task returns than template-only dialogue, even with observable cognitive engagement held constant. Therefore, starters are better designed as an entry point into student-generated dialogue than as a substitute for it. For example, a starter can invite students to identify the difficulty, explain what they have already considered, or specify the aspect of their work on which they need feedback before the companion responds substantively. Such designs retain the accessibility of predefined support while creating space for students to contribute their own messages.

Companion design should also support cognitive engagement and subsequent task action. Constructive and Interactive dialogue were associated with higher final artefact quality, and these modes involved reasoning, evaluation, critique, and sustained uptake. AI companions can invite these forms of student contribution by asking for explanations, requesting justification for a proposed answer, prompting comparison between alternatives, or inviting students to evaluate AI-generated feedback before acting on it. Socratic questioning that withholds direct answers in favour of guiding learners towards their own reasoning offers one concrete mechanism for eliciting such engagement \citep{favero2025socratic, xi2026socratic}. What happens after the dialogue warrants equal design attention: interfaces can connect AI responses to the specific task actions they bear on, such as revising an option or reconsidering feedback, so that returning to the task is itself supported rather than left to the student.

For evaluation, prompt counts and time spent with AI cannot distinguish these properties. Instead, learning analytics for task-embedded AI can record whether students contributed their own messages or used only templates, the cognitive engagement evident in their dialogue, where the interaction occurred within the task process, and what followed it. Kept separate, these indicators reveal how AI is being used for answer seeking, explanation, evaluation, or revision, uses that a single composite metric would obscure. In practice settings, the same logic applies to outcomes: continuation and subsequent correctness should be tracked as separate indicators rather than read from a single measure of post-AI activity. These findings establish empirically grounded priorities for design and analytics, which experimental studies can now test for their effects on engagement and outcomes.

\section{Limitations and Future Research}
This study draws on authentic task-embedded AI use, and several boundaries of this evidence point to directions for further research. First, students chose when and how to seek AI assistance, so the relationships between dialogue characteristics, task processes, and outcomes are associational. Experimental manipulation of prompt suggestions, companion prompts, and the timing of AI support can test how specific design features shape cognitive engagement and subsequent task activity. Second, engagement composition and transition structures were compared at the population level and may partly reflect students' self-selection into companion use and into composing their own messages, as well as the workflows the platform made available, rather than change within individual students. Tracking repeated tasks from the same students would address this within-person question. Third, artefact quality and practice performance were measured during AI-supported coursework, so the findings concern the quality of assisted task completion rather than gains in independent knowledge or skills. Pre-AI and post-AI performance measures would show whether the engagement patterns identified here are associated with sustained learning. Fourth, although both primary contrasts remained significant after Holm correction and the engagement-quality association was consistent in direction across the estimable sensitivity analyses, several comparison groups were small, limiting the precision of secondary, cross-path, and within-student estimates. Larger samples would allow these estimates to be tested more rigorously. Fifth, platform logs captured AI use within RiPPLE but not assistance from external tools such as ChatGPT or Claude, which students commonly use alongside learning platforms \citep{park2024promise_peril_chatgpt_he}. Finally, process traces identified where students moved after AI dialogue and how often they returned to the task, but not how AI feedback was taken up in revisions. Linking revision histories directly to AI responses would reveal which suggestions students adopt, reject, or modify, and how these changes shape the developing artefact.

\section{Conclusion}
This study examined how students cognitively engaged with task-embedded AI, how dialogue was situated within learning workflows, and how these characteristics related to post-dialogue activity and task outcomes. Because interaction frequency reveals little about what students accomplish with task-embedded AI, the study connected the cognitive engagement in student-AI dialogue with the task processes in which it was embedded and the outcomes that followed, providing converging evidence that post-dialogue activity and task outcomes are distinct aspects of AI-supported learning. Cognitive engagement and the position of AI dialogue both differed across workflows. In resource creation, student-composed messages were associated with behavioural re-engagement, whereas deeper cognitive engagement was associated with higher final artefact quality, an association that return frequency did not account for. In practice tasks, AI dialogue after an incorrect attempt was associated with lower continuation but higher next-attempt correctness among students who continued. Thus, for design and learning analytics, message source, cognitive engagement, workflow position, and post-dialogue activity should be recorded as separate indicators. Accordingly, evaluating AI in education requires moving beyond how often students use AI towards the cognitive engagement evident in their dialogue and what they do next.

%TC:ignore
% Statements
\section*{Author Contributions}
\textbf{Xiaoyu Zhuang:} Conceptualization, Data Curation, Formal Analysis, Investigation, Methodology, Software, Validation, Visualization, Writing – Original Draft Preparation, Writing – Review \& Editing.

\textbf{Aneesha Bakharia:} Conceptualization, Investigation, Methodology, Project Administration, Software, Supervision, Writing – Review \& Editing.

\textbf{Dragan Gašević:} Conceptualization, Supervision, Writing – Review \& Editing.

\textbf{Hassan Khosravi:} Conceptualization, Methodology, Project Administration, Resources, Supervision, Visualization, Writing – Review \& Editing.

\section*{Acknowledgements}
We thank Dr Jonathan Brazil for his assistance with the ICAP coding of student-AI conversation threads.

\section*{Funding}
This research was supported by a PhD Research Project Scholarship from The University of Queensland.

\section*{Ethics Statement}
This study was approved by The University of Queensland Human Research Ethics Committees (HRECs) (Approval No. 2023/HE001453).

\section*{Conflicts of Interest}
The authors declare no conflicts of interest.

\section*{Data Availability Statement}
The data that support the findings of this study are available upon reasonable request, subject to ethical and privacy restrictions.

% References
\printbibliography

% Appendices
\appendix
\section{Workflow State Definitions}
\label{app:state_definitions}
Full state definitions for the workflow-specific sequences.

\begin{table}[H]
\centering
\caption{States for the Creation workflow.}
\label{tab:creation_states}
\begin{tabularx}{\linewidth}{p{0.18\linewidth} X}
\hline
\textbf{State} & \textbf{Description} \\
\hline
\texttt{START\_SESSION} &
Synthetic state indicating the beginning of a Creation sequence. \\

\texttt{OPTIONS} &
Creation or modification of response options or distractors associated with the resource. \\

\texttt{INITIAL\_CONTENT} &
The opening uninterrupted phase in which the student creates initial resource content (excluding answer options or distractors) from blank fields. \\

\texttt{CONTENT\_EDITING} &
Modification of resource content (excluding answer options or distractors) after the opening content-creation phase. \\

\texttt{SELF\_EVALUATION} &
Completion or revision of self-evaluation information about the created resource. \\

\texttt{AI\_CHAT} &
A student's message sent to the Crea AI companion, irrespective of its ICAP label. \\

\texttt{AI\_FEEDBACK} &
Receipt, selection, or evaluation of AI-generated feedback on the resource. \\

\texttt{SUBMIT} &
Submission of the created learning resource. \\

\texttt{END\_SESSION} &
Synthetic state indicating the end of a Creation sequence. \\
\hline
\end{tabularx}
\end{table}

\begin{table}[H]
\centering
\caption{States for the Moderation workflow.}
\label{tab:moderation_states}
\begin{tabularx}{\linewidth}{p{0.25\linewidth} X}
\hline
\textbf{State} & \textbf{Description} \\
\hline
\texttt{START\_SESSION} &
Synthetic state indicating the beginning of a Moderation sequence. \\

\texttt{MODERATION\_FEEDBACK} &
Creation or revision of evaluative feedback on a learning resource. \\

\texttt{MODERATION\_SUGGESTION} &
Creation or revision of a suggested improvement to the resource. \\

\texttt{MODERATION\_COMMENT} &
Creation or revision of an additional moderation comment. \\

\texttt{AI\_CHAT} &
A student's message sent to the Modi AI companion, irrespective of its ICAP label. \\

\texttt{MODERATION\_SUBMIT} &
Submission of the completed moderation response for the focal learning resource. \\

\texttt{END\_SESSION} &
Synthetic state indicating the end of a Moderation sequence. \\
\hline
\end{tabularx}
\end{table}

\begin{table}[H]
\centering
\caption{States for the Practice workflow.}
\label{tab:practice_states}
\begin{tabularx}{\linewidth}{p{0.22\linewidth} X}
\hline
\textbf{State} & \textbf{Description} \\
\hline
\texttt{START\_SESSION} &
Synthetic state indicating the beginning of a Practice sequence. \\

\texttt{PRACTICE\_INCORRECT} &
An incorrect answer submitted by the student during practice. Consecutive occurrences are retained as distinct attempts. \\

\texttt{AI\_CHAT} &
A student's message sent to the Praxie AI companion, irrespective of its ICAP label. \\

\texttt{PRACTICE\_CORRECT} &
A correct answer submitted by the student during practice. Consecutive occurrences are retained as distinct attempts. \\

\texttt{END\_SESSION} &
Synthetic state indicating the end of a Practice sequence. \\
\hline
\end{tabularx}
\end{table}

\section{Model Specifications}
\label{app:model_specs}
Equations and specifications for the models used in the outcome analysis.

\subsection*{Behavioural re-engagement}
For artefact $a$, count of dialogue-to-task transitions was

\begin{equation}
T_a=
\sum_t
\mathbf{1}\!\left[
S_t=\texttt{AI\_CHAT},
\;
S_{t+1}\in
\{\texttt{CONTENT\_EDITING},\texttt{OPTIONS},\texttt{SELF\_EVALUATION}\}
\right],
\label{eq:reengagement}
\end{equation}

where $S_t$ denotes the current state in Table~\ref{tab:creation_states} and $S_{t+1}$ denotes the next state.

Re-engagement was estimated using Gaussian GEE

\begin{equation}
E(T_a)
=
\alpha_T+\delta_TD_a+\boldsymbol{\gamma}_T^{\mathsf T}\mathbf{X}_a,
\label{eq:gee}
\end{equation}

where $D_a$ denotes the dialogue contrast and $\mathbf{X}_a$ contains prior mean peer rating, unavailable rating history, log prior rated-artefact count, submission order, and offering fixed effects.

\subsection*{Artefact quality}
For rating $Y_{ar}\in\{1,2,3,4,5\}$ of artefact $a$ by reviewer $r$, quality was modelled with cumulative-logit regression

\begin{equation}
\operatorname{logit}P(Y_{ar}\leq k)
=
\tau_k-
\left(
\beta D_a
+\boldsymbol{\gamma}_Y^{\mathsf T}\mathbf{X}_a
+\lambda S^{(-ar)}_{ar}
+\psi M_{ar}
+\boldsymbol{\theta}^{\mathsf T}\mathbf{Z}_a
\right),
\label{eq:cumlogit}
\end{equation}

where $S^{(-ar)}_{ar}$ is the reviewer's leave-one-out mean rating and $M_{ar}$ indicates Modi companion use before the focal rating. For reviewers with only one rating on a dimension, the sample mean for that dimension was used.

The post-dialogue process vector was

\begin{equation}
\mathbf{Z}_a=
\left[
z(T_a),\;
Q_a,\;
z(A_a),\;
z(C_a)
\right]^{\mathsf T},
\label{eq:zvector}
\end{equation}

where $Q_a$ indicates premature closure, $A_a$ is the log-transformed elapsed time from the first AI event to submission, $C_a$ is the number of pre-submission AI-chat events, and $z(\cdot)$ denotes standardisation within the analytic sample, not applied to $Q_a$. The total model fixed $\boldsymbol{\theta}=\mathbf{0}$; the process-adjusted model estimated $\boldsymbol{\theta}$ freely. Because $\mathbf{Z}_a$ was measured after dialogue, the latter was interpreted as conditional adjustment rather than mediation. Inference used author\,$\times$\,reviewer two-way cluster-robust covariance.

Adjusted expected ratings from the total model were

\begin{equation}
\widehat{\mu}(d)
=
\frac{1}{N_R}
\sum_{(a,r)\in\mathcal{R}}
\sum_{k=1}^{5}
k\,\widehat{P}(Y_{ar}=k\mid D_a=d,\mathbf{X}_a,S^{(-ar)}_{ar},M_{ar}),
\label{eq:standardisation}
\end{equation}

where $\mathcal{R}$ is the set of reviewer-level records and $N_R=|\mathcal{R}|$. The proportional-odds check allowed $\beta$ to vary across thresholds, retaining only thresholds with at least five observations in every exposure-by-dichotomised-outcome cell. Peer-rating agreement used linear-weighted Gwet's AC2 with 95\% percentile CIs from 1,000 artefact-level bootstrap resamples.

\subsection*{Practice outcomes}
For landmark-eligible incorrect attempt $o$,

\begin{equation}
\operatorname{logit}P(Y_o=1)
=
\alpha_P+\delta_PE_o+\boldsymbol{\gamma}_P^{\mathsf T}\mathbf{W}_o,
\label{eq:practice}
\end{equation}

where $E_o$ indicates AI dialogue within 30 seconds and $\mathbf{W}_o$ contains learner ability, resource difficulty, mastery rating, attempt position, preceding incorrect attempts, and offering fixed effects, with offerings contributing fewer than five opportunities pooled into a single category. Outcomes were observed over $30<t\leq630$ seconds. Inference used student\,$\times$\,resource two-way cluster-robust covariance.

\begin{table}[H]
\centering
\caption{Practice outcome definitions.}
\label{tab:practice_outcomes}
\begin{tabular}{p{0.25\linewidth} p{0.40\linewidth} p{0.25\linewidth}}
\toprule
\textbf{Outcome} & \textbf{$Y_o=1$ when} & \textbf{Sample} \\
\midrule
Continuation
& A further attempt occurs.
& All opportunities. \\
Conditional correctness
& The next attempt is correct.
& Continuing opportunities. \\
Joint correct recovery
& A further attempt occurs and is correct.
& All opportunities. \\
\bottomrule
\end{tabular}
\end{table}

\subsection*{Sensitivity analyses}

\begin{table}[H]
\centering
\caption{Sensitivity analyses.}
\label{tab:sensitivity}
\begin{tabular}{p{0.25\linewidth} p{0.65\linewidth}}
\toprule
\textbf{Check} & \textbf{Specification} \\
\midrule
Artefact-level means
& Mean rating per artefact, weighting artefacts equally. \\
Moderator AI use
& Ratings preceded by Modi use excluded. \\
Prior rating history
& Authors required at least one previously rated artefact. \\
Leave-one-offering-out
& Each contributing offering omitted in turn. \\
Landmark window
& Practice landmark varied to 15 and 60 seconds. \\
Within-student
& Mundlak within-between model among students observed under both exposure conditions, where estimable. \\
\bottomrule
\end{tabular}
\end{table}

\section{Representative ICAP-Coded Conversation Threads}
\label{app:icap_examples}
Examples of ICAP-coded anonymised conversation threads and coding rationales.

\subsection*{Off-task}

\begin{conversation}
\textbf{Student:} ``The page is not loading to the next question.''

\vspace{0.5\baselineskip}

\textbf{AI:} ``What if you try refreshing the page or checking your internet connection? Sometimes a simple reload fixes loading issues. Have you tried opening the page in a different browser? Could clearing your cache or cookies help? Is there an option to access the next question through a different link or menu? Let's explore it together - step by step.''
\end{conversation}

\noindent\textit{Coding rationale.} The student's only message concerns a platform/interface loading problem (``page is not loading to the next question''), which is a platform mechanics issue rather than academic engagement.

\subsection*{Passive}

\begin{conversation}
\textbf{Student:} ``What should I write in the explanation boxes?''

\vspace{0.5\baselineskip}

\textbf{AI:} ``Great question! In the explanation boxes for each part of your MCQ, you should write clear and concise reasons why each answer choice is correct or incorrect. For the correct answer, explain why it's the best choice, referencing key concepts from the topic. For the distractors, briefly explain why they are not correct, which helps learners understand common misconceptions or errors. This approach strengthens the learning experience and deepens understanding. If you haven't selected a topic yet, I recommend choosing one first so your explanations can be aligned with the relevant content.''
\end{conversation}

\noindent\textit{Coding rationale.} The student asks a generic on-task resource-authoring question about what to put in explanation boxes, without providing specific academic content, an attempt, or reasoning. This fits Passive engagement rather than Active or higher.

\subsection*{Active}

\begin{conversation}
\textbf{Student:} ``How does increased spending influence economic growth?''

\vspace{0.5\baselineskip}

\textbf{AI:} ``Increased government spending puts more money into the economy, raising demand for goods and services. This encourages businesses to produce more and hire more workers. Such activity usually boosts economic growth. How might this relate to the government's investments in Medicare services in the budget?''
\end{conversation}

\noindent\textit{Coding rationale.} The student's own message asks a specific conceptual economics question about how increased spending influences economic growth, but it does not include their own reasoning, explanation, critique, or reflection. Under the provided ICAP guidance, specific content-related questions without self-reasoning are Active rather than Constructive.

\subsection*{Constructive}

\begin{conversation}
\textbf{Student:} ``I am looking to write my resource on IOT and wearable tech. The insight that I've gained is that IOT and wearable tech possess a rising breach in privacy. My position is to counter the increased breach of privacy with restrictions or highly visible clauses in agreements so that people understand what is being monitored. How can I structure my resource?''

\vspace{0.5\baselineskip}

\textbf{AI:} ``Nice focus - that's a clear, class-relevant insight. Here's a tight structure you can use to turn your idea into the required note without me writing the content for you. Use only your material when filling each part.
\begin{itemize}
    \item Title: short, specific (e.g., `Wearables, IoT, and the Privacy Gap')
    \item Section 1 (Opening insight): one sentence stating the insight you gained from the guest speaker/debate (privacy breaches are rising with wearables).
    \item Section 1 Explanation (Context \& evidence): 2-3 short bullets citing the guest speaker or debate example(s) that showed how data is collected or mishandled. Name the source (speaker/debate point).
    \item Section 2 (Position): one sentence stating your stance (require restrictions / highly visible clauses).
    \item Section 2 Explanation (Rationale \& mechanism): 3 bullets linking your position to specific harms and listing concrete ways visibility/restrictions would help (e.g., clear monitoring clauses, opt-in dashboards, simplified consent). Use only concepts you already mentioned.
    \item Section 3 (Class advancement): 1-2 sentences explaining how this insight advances class understanding - what classmates should now consider or discuss.
    \item Closing line: one sentence calling for action or further debate (e.g., evaluate current agreements for transparency).
\end{itemize}

A few micro-guidance points while you fill the parts:
\begin{itemize}
    \item In the Explanation bullets, quote or paraphrase the speaker exactly when possible (keep citations short).
    \item Avoid introducing new technical details you haven't discussed; stay focused on privacy and consent clarity.
    \item Keep each part concise so the note meets the assessment prompt.
\end{itemize}

Would you like me to check a draft if you paste your filled sections?''
\end{conversation}

\noindent\textit{Coding rationale.} The student articulates an original insight and position, then explains the rationale for visible agreement clauses - so people understand what is monitored. This shows student-generated interpretation and reasoning, but there is no sustained back-and-forth to support Interactive.

\subsection*{Interactive}

\begin{conversation}
\textbf{Student:} ``Please assist me by providing a review of my resource in its current state.''

\vspace{0.5\baselineskip}

\textbf{AI:} ``Thanks - nice work submitting this MCQ on the `Error Prevention' heuristic.

Strengths:
\begin{itemize}
    \item Clear Question Stem: The question directly asks which option best matches the Error Prevention heuristic, so learners know exactly what to evaluate.
    \item Single correct answer chosen: Option B is correctly marked as the one best matching Error Prevention, meeting the single-correct rule.
    \item Good distractors with explanations: Options A, C, and D are plausible but clearly wrong for the heuristic; each has a targeted explanation that distinguishes it from the correct heuristic.
\end{itemize}

Improvements:
\begin{itemize}
    \item Option A (`Ensure users are never informed of errors'): Issue - phrasing is slightly awkward and could be interpreted as ambiguous; Revision - rephrase this existing option text to a clearer negative statement (use the same idea but with more direct wording so learners immediately see it as bad design).
    \item Option D (`Use good coding practices to make sure users cannot break the system'): Issue - may blur system robustness with user-facing error prevention; Revision - clarify this existing option to emphasize system-level coding practices (so it's clearly about backend robustness rather than user-error prevention).
\end{itemize}

Conclusion: This MCQ is ready to use as-is for assessment; the two suggested wording clarifications are optional refinements to reduce any momentary ambiguity. Great idea! Let's craft it into a platform-worthy resource.''

\vspace{0.5\baselineskip}

\textbf{Student:} ``I have actioned on the 2 revisions, did i make it worse or better?''

\vspace{0.5\baselineskip}

\textbf{AI:} ``Nice - thanks for updating the two options. I checked the current MCQ (Question Stem + Options A - D) against the Error Prevention heuristic and your earlier feedback. Your revisions improved clarity overall:
\begin{itemize}
    \item Option A (`Do not inform users of errors that occur'): The phrasing now clearly signals a poor practice and ties to Visibility of System Status - keeping it a clearly incorrect distractor.
    \item Option D (`Ensure system robustness so that users cannot cause system errors'): The wording now distinguishes backend robustness from user-facing error prevention, so it's plausibly wrong for this heuristic.
\end{itemize}

One small remaining nitpick to consider (optional): Option A's negative framing is clear but blunt; if you prefer softer language while keeping it incorrect, reword the existing sentence rather than replacing its intent.''

\vspace{0.5\baselineskip}

\textbf{Student:} ``OK is this better? I toned down the negatives and made it seem more plausible to be the correct option.''

\vspace{0.5\baselineskip}

\textbf{AI:} ``Nice revision - your toning-down improves educational value while keeping distractors plausible.

Specific checks:
\begin{itemize}
    \item Question Stem: Clear and aligned with `Error Prevention.' No change needed.
    \item Option A (`Hide errors from users to ensure a seamless experience'): Now softer but still clearly wrong; the explanation correctly ties it to Visibility of System Status. Keep as-is.
    \item Option B (`Warn users before they follow through with an action that might be a mistake'): Still the single correct choice; the explanation correctly links warnings to preventing user errors.
    \item Option C (`Make it easy for users to recover from errors'): Correctly marked as incorrect for this heuristic; the explanation distinguishes recovery vs prevention - keep.
    \item Option D (`Ensure system robustness so that users cannot cause system errors'): Wording now reads as plausible but remains distinct from user-focused prevention; the explanation correctly contrasts system robustness with user-facing prevention.
\end{itemize}

Final judgment: This MCQ is better - clearer, with more plausible distractors, while still satisfying the single-correct rule. It's ready to submit. Great idea! Let's craft it into a platform-worthy resource.''
\end{conversation}

\noindent\textit{Coding rationale.} The student repeatedly applies the AI's feedback and iteratively refines the resource across turns, first actioning two revisions and then adjusting negative wording to make distractors more plausible. This sustained uptake and design refinement constitutes co-construction rather than a one-time review request.

%TC:endignore

\end{document}